\documentclass[reprint,twocolumn,superscriptaddress,showkeys,
nofootinbib,notitlepage,amsmath,amssymb,floatfix]{revtex4-1}

\usepackage{latexsym,amsmath,amssymb,lmodern,float,url}
\usepackage{natbib}
\usepackage{color}
\usepackage{multirow}
\usepackage{bm}
\usepackage{array}

\usepackage{mathtools}
\usepackage{physics}

\def\de{\delta^{\vphantom{1}}}
\def\bde{{\bar\delta}}
\def\ccqq{{c\bar{c}q\bar{q}^\prime}}
\def\bbqq{{b\bar{b}q\bar{q}^\prime}}

\def\qq{{q\bar q}}
\def\QQ{{Q\bar Q}}
\def\cc{{c\bar c}}

\def\bt{{\bar\theta}}

\def\h3{{\displaystyle{\frac 3 2}}}

\begin{document}
\title{Radiative Transitions of Charmoniumlike Exotics in the Dynamical Diquark Model}
\author{Justin M. Gens}
\email{jgens@asu.edu}
\author{Jesse F. Giron}
\email{jfgiron@asu.edu}
\author{Richard F. Lebed}
\email{Richard.Lebed@asu.edu}
\affiliation{Department of Physics, Arizona State University, Tempe,
AZ 85287, USA}
\date{February, 2021}

\begin{abstract}
Using the dynamical diquark model, we calculate the electric-dipole
radiative decay widths to $X(3872)$ of the lightest negative-parity
exotic candidates, including the four $I=0$, $J^{PC} \! = \! 1^{--}$
(``$Y$'') states.   The $O$(100--1000~keV) values obtained test the
hypothesis of a common substructure shared by all of these states.
We also calculate the magnetic-dipole radiative decay width for
$Z_c(4020)^0 \! \to \! \gamma X(3872)$, and find it to be rather
smaller ($<$~10~keV) than its predicted value in molecular models.
\end{abstract}

\keywords{Exotic hadrons, radiative transitions, diquarks}
\maketitle
\section{Introduction}
The number of new heavy-quark exotic-hadron candidates, presumptive
tetraquark and pentaquark states, increases every year.  In the past
18 years, over 40 candidates have been observed at multiple
facilities and their hosted experiments.  However, no single
theoretical picture to describe the structure of these states has
emerged as an undisputed favorite.  Both the broad scope of
experimental results and competing theoretical interpretations have
been reviewed by many in recent years~\cite{Lebed:2016hpi,
Chen:2016qju,Hosaka:2016pey,Esposito:2016noz,Guo:2017jvc,Ali:2017jda,
Olsen:2017bmm,Karliner:2017qhf,Yuan:2018inv,Liu:2019zoy,
Brambilla:2019esw}.

Among these competing physical approaches, the dynamical diquark
picture~\cite{Brodsky:2014xia} was developed to provide a mechanism
through which diquark~($\de$)-antidiquark~($\bde$) states could
persist long enough to be identified as such experimentally.
Diquarks are formed through the attractive channels ${3} \otimes
\bf{3} \! \to \! \bar{\bf{3}}$ [$\delta \! \equiv \!
(Q q)_{\bar{\bf 3}}$] and $\bar{\bf 3} \otimes \bar{\bf 3} \! \to \!
{\bf 3}$ [$\bde \! \equiv \! (\bar Q {\bar q}^\prime)_{\bf 3}$]
between color-triplet quarks.  In this physical picture, the heavy
quark $Q$ must first be created in closer spatial proximity to a
light quark $q$ than to a light antiquark $\bar q^\prime$ (and vice
versa for $\bar Q$).  This initial configuration provides an
opportunity for the formation of fairly compact $\de$ and $\bde$
quasiparticles, in distinction to an initial state in which the
strongly attractive $\bf{3} \otimes \bar{\bf{3}} \! \to \! \bf{1}$
coupling immediately leads to $(Q\bar q^\prime)(\bar Q q)$ meson
pairs.  Second, the large energy release of the production process
(from a heavy-hadron decay or in a collider event) drives apart the
$\de$-$\bde$ pair before immediate recombination into a meson pair
can occur, creating an observable resonance.  A similar mechanism
extends the picture to pentaquark formation~\cite{Lebed:2015tna},
by means of using color-triplet ``antitriquarks'' $\bt \! \equiv \! [\bar
Q_{\bar{\bf 3}} (q_1 q_2)_{\bar{\bf 3}}]^{\vphantom{(}}_{\bf 3}$.

This physical picture was subsequently developed into the dynamical
diquark {\em model}~\cite{Lebed:2017min}: The separated $\de$-$\bde$
pair is connected by a color flux tube, whose quantized states are
best described in terms of the potentials computed using the
Born-Oppenheimer (BO) approximation.  These are the same potentials
as appear in QCD lattice gauge-theory simulations that predict the
spectrum of heavy-quarkonium hybrid mesons~\cite{Juge:1997nc,
Juge:1999ie,Juge:2002br,Morningstar:2019,Capitani:2018rox}.  The BO
potentials are introduced into coupled Schr\"{o}dinger equations that
are solved numerically in order to produce predictions for the
$\de$-$\bde$ spectrum, as shown in Ref.~\cite{Giron:2019bcs}.  As one
of the primary results of that work, all the observed exotic
candidates are shown to be accommodated within the ground-state
BO potential $\Sigma^+_g$, with the specific multiplets in order of
increasing average mass being $1S$, $1P$, $2S$, $1D$, and $2P$.  A
full summary of the BO potential notation is presented in
Ref.~\cite{Lebed:2017min}.

The mass spectrum and preferred decay modes (organized by eigenstates
of heavy-quark spin) of the 6 isosinglets and 6 isotriplets
comprising the $\ccqq$ positive-parity $\Sigma^+_g(1S)$ multiplet
(where $q,q^\prime \in \{ u, d \}$) were studied in
Ref.~\cite{Giron:2019cfc}.  This was the first work to differentiate
$I \! = \! 0$ and $I \! = \! 1$ states in a diquark model.  The
specific model of Ref.~\cite{Giron:2019cfc} naturally produces
scenarios in which $X(3872)$ is the lightest $\Sigma^+_g(1S)$ state,
and moreover predicts that the lighter of the two $I \! = \! 1$,
$J^{PC} \! = \! 1^{+-}$ states in $\Sigma^+_g(1S)$ [$Z_c(3900)$]
naturally decays almost exclusively to $J/\psi$ and the heavier one
[$Z_c(4020)$] to $h_c$, as is observed.  The model of
Ref.~\cite{Giron:2019cfc} uses a 3-parameter Hamiltonian consisting
of a common multiplet mass, an internal diquark-spin coupling, and a
long-distance isospin- and spin-dependent coupling (analogous to
$\pi$ exchange) between the light quark $q$ in $\de$ and light
antiquark $\bar q^\prime$ in $\bde$.  Similar conclusions using QCD
sum rules have been obtained in Ref.~\cite{Ghalenovi:2020zen}.

The dynamical diquark model was developed further through the
corresponding analysis~\cite{Giron:2020fvd} of the negative-parity
$\ccqq$ $\Sigma^+_g(1P)$ multiplet and its 28 constituent
isomultiplets (14 isosinglets and 14 isotriplets), which includes
precisely four $Y$ ($I \! = \! 0$, $J^{PC} \! = \!  1^{--}$) states.
In this case, the simplest model has 5 parameters: the 3 listed
above, plus spin-orbit and tensor terms. An earlier diquark analysis
using a similar Hamiltonian, but not including isospin dependence,
appears in Ref.~\cite{Ali:2017wsf}.

The success of Ref.~\cite{Giron:2019bcs} in predicting the correct
mass splittings between the observed bands ($1S,1P,2S$) of exotic
hadrons, and Ref.~\cite{Giron:2019cfc} in effectively representing
the fine structure within the lowest multiplets [especially
$\Sigma^+_g(1S)$] provides strong {\it a posteriori\/} support for
the applicability of the dynamical diquark model.  In particular, one
may certainly question whether treating the exotics as quasi-two-body
states within a BO approximation, rather than including full 4- (or
5-) body interactions to represent the internal evolution of the
quasiparticles, is sensible.  However, while such effects are
undoubtedly present at some level, the current experimental
evidence appears to support the presence of a scale separation that
allows the quasiparticles to be treated identifiable subunits within
the hadrons.  As an example, Ref.~\cite{Giron:2019bcs} showed in
numerical simulations that the diquarks need not be pointlike
particles, but could have substantial spatial extent (characteristic
radii as large as $0.4$~fm) before the full hadron mass spectrum
changes significantly.

An analysis within this model of the 12 isomultiplets comprising the
$\bbqq$ $\Sigma_g^+(1S)$ multiplet and the 6 states of the
$\cc s\bar{s}$ $\Sigma_g^+(1S)$ multiplet appears in
Ref.~\cite{Giron:2020qpb}. By using only experimental inputs for the
states $Z_b(10610)$ and $Z_b(10650)$, which includes their masses and
relative probability of decay into $h_b$ versus $\Upsilon$ states, the
entire $\bbqq$ mass spectrum is predicted.  In particular, the mass
of the bottom analogue to $X(3872)$ is highly constrained
($\approx \! 10600$~MeV), and the lightest $\bbqq$ state
($I \! = \! 0$, $J^{PC} \! = \! 0^{++}$) lies only a few MeV above
the $B\bar{B}$ threshold.  Furthermore, starting with the assumption
that $X(3915)$ is the lowest lying $\cc s\bar{s}$
state~\cite{Lebed:2016yvr} and $Y(4140)$ is the sole $J^{PC} \! = \!
1^{++}$ $\cc s\bar{s}$ state in $\Sigma_g^+(1S)$, the remaining 4
masses in the multiplet are predicted.  Emerging naturally in the
spectrum is $X(4350)$, a $J/\psi$-$\phi$ resonance seen by
Belle~\cite{Shen:2009vs}, while $Y(4626)$ and $X(4700)$ are found to
fit well within the $\Sigma_g^+(1P)$ and $\Sigma_g^+(2S)$
$\cc s\bar{s}$ multiplets, respectively.

The dynamical diquark model has also recently been extended to the
case in which the light quarks $q$ are replaced with heavy quarks $Q$
to produce fully heavy tetraquark states $Q_1 \overline Q_2 Q_3
\overline Q_4$, where $Q_i \! = \! c$ or $b$.  Sparked by the recent
LHCb report of at least one di-$J/\psi$ resonance near $6900$
MeV~\cite{Aaij:2020fnh}, Ref.~\cite{Giron:2020wpx} determined the
spectrum of $c\bar{c}c\bar{c}$ states in the dynamical diquark model.
In this system, the minimal model predicts each $S$-wave multiplet to
consist of 3 degenerate states ($J^{PC} \! = \! 0^{++},\,1^{+-},\,
2^{++}$) and 7 $P$-wave states.  $X(6900)$ was found to fit most
naturally as a $\Sigma_g^+(2S)$ state, with other structures in the
measured di-$J/\psi$ spectrum appearing to match $C \! = \! +$
members of the $\Sigma^+_g(1P)$ multiplet.

In this paper we use the dynamical diquark model to predict
radiative transitions between exotic states.  So far, very few
theoretical papers have investigated exotic-to-exotic transitions
(and of these papers, only diquark models have been
considered~\cite{Maiani:2014aja,Chen:2015dig}).  One of the
distinctive features of the $P$-wave study in
Ref.~\cite{Giron:2020fvd} is the direct calculation of decay
probabilities to eigenstates of heavy-quark spin.  Indeed,
Ref.~\cite{Giron:2020fvd} uses the heavy quark-spin content of states
as the main criterion for associating observed resonances with
particular states in the $\Sigma^+_g(1P)$ multiplet, and identifies
using likelihood fits two particularly plausible assignments for the
states.  Using the same decay probabilities, we calculate here
the transition amplitudes for $\Sigma^+_g(1P) \! \to \! \gamma
\Sigma^+_g(1S)$.  We directly adapt the well-known expression for
electric dipole (E1) radiative transitions used to great effect for
conventional quarkonium.  Since the E1 transition formula depends
sensitively upon the initial and final wave functions, a comparison
between our predictions and data provides an important test of the
hypothesis that the purported $\Sigma^+_g(1P)$ and $\Sigma^+_g(1S)$
states, such as in $Y(4220) \! \to \! \gamma X(3872)$, truly share a
common structure.  The corresponding magnetic dipole (M1) expression
within this model is also presented, in anticipation of the
observation of relevant transitions such as $\Sigma^+_g(2S) \! \to \!
\gamma \Sigma^+_g(1S)$, or even between two $\Sigma^+_g(1S)$ states
such as $Z_c(4020)^0 \! \to \! \gamma X(3872)$.

This paper is organized as follows: In Sec.~\ref{sec:Expt} we review
the current experimental data on transitions between $\ccqq$ states.
Section~\ref{sec:pwave_review} reprises the relevant phenomenological
aspects of Ref.~\cite{Giron:2020fvd}.  In Sec.~\ref{sec:analysis} we
calculate the decay widths and decay probabilities for
exotic-to-exotic radiative transitions and focus upon two of the more
probable $P$-wave state assignments in Ref.~\cite{Giron:2020fvd}.
We conclude in Sec.~\ref{sec:conclusions}.

\section{Experimental Review of Exotic-to-Exotic Transitions}
\label{sec:Expt}

Although the number of exotic-candidate discoveries continues to
increase at a remarkable pace, only a handful of exotic-to-exotic
decays have been observed to date, through
radiative~\cite{Ablikim:2013dyn,Ablikim:2019zio} and
pionic~\cite{Ablikim:2013mio,Liu:2013dau,BESIII:2020pov} transitions.

Considering first the radiative decays that form the topic of this
work, thus far only E1 transitions (as indicated by changing parity
$\Delta P \! = \! -$) have been observed in two states at BESIII, the
$J^{PC} \! = \! 1^{--}$ $Y(4260)$~\cite{Ablikim:2013dyn} and
$Y(4220)$~\cite{Ablikim:2019zio}, both seen to decay to a photon and
the $J^{PC} \! = \! 1^{++}$ $X(3872)$.  Indeed, an increasing amount
of evidence from BESIII ({\it e.g.}, in Ref.~\cite{Ablikim:2016qzw})
suggests that the well-known $Y(4260)$ is actually a collection of
resonances, of which $Y(4220)$ is just one component.
Observed exotic-to-conventional radiative transitions are also rather
few in number, due to the large decay widths of exotics that follows
from the dominance of their strong decay modes.  To date, only
$X(3872) \! \to \! \gamma J/\psi$ and $\gamma \psi(2S)$, also both E1
transitions, have definitely been seen ({\it e.g.}, in
Ref.~\cite{Aaij:2014ala}).  BESIII has also recently announced an 
interesting negative result~\cite{Ablikim:2021rba}, an upper limit
for $Z_c(4020)^0 \, (J^{PC} \! = \! 1^{+-}) \! \to \! \gamma
X(3872)$.  Indeed, to date no M1 radiative decay
($\Delta P \! = \! +$) of any exotic candidate has yet been seen at
any experiment.

As for pionic transitions, both BESIII~\cite{Ablikim:2013mio} and
Belle~\cite{Liu:2013dau} have observed (indeed, discovered)
$Z_c(3900)^\pm$ through $Y(4260) \! \to \! \pi^+ \pi^- J/\psi$, and
BESIII recently observed $Z_c(3900)^0$ via $Y(4220) \! \to \! \pi^0
\pi^0 J/\psi$~\cite{BESIII:2020pov}.  Assuming just a similarity of
hadronic structure between various exotic candidates, one may expect
several more exotic-to-exotic pionic (or other light-meson)
transitions to be observed in the future.  An essential criterion for
how such transitions may best be studied relies on the size of the
pion momentum $p_\pi$ in such processes; for example, in the decays
listed above, $p_\pi \! \approx \! 300$~MeV\@.  Processes with
smaller $p_\pi$ values may be reliably studied using conventional
chiral perturbation theory, while studies of processes with larger
$p_\pi$ values require modifications to the perturbative calculation
to improve their convergence.  Since the methods associated with
radiative transitions (particularly E1 transitions) present fewer
computational ambiguities, we defer a study of exotic-to-exotic
pionic transitions for future work.

The expressions for E1 and M1 transition widths used below
[Eqs.~(\ref{eqn:e1width}) and (\ref{eqn:m1width}), respectively] are
almost identical to the forms derived in standard quantum mechanics
textbooks.  As such, they are manifestly nonrelativistic, and
furthermore are developed using the photon long-wavelength
approximation, $\exp ( i \bm{k} \cdot \bm{r} ) \! \to \! 1$.
Nevertheless, the expressions can also be derived directly from the
fundamental Lagrangian $j_\mu A^\mu$ couplings of the electromagnetic
current $j^\mu$ of charged quarks to the photon field $A^\mu$ (see,
{\it e.g.}, Ref.~\cite{Brambilla:2004wf}).  In
Sec.~\ref{sec:pwave_review} we discuss the effect of including
certain corrections to the textbook expressions.

In the dynamical diquark model, all states in the multiplet
$\Sigma^+_g(1S) [(1P)]$ have $P \! = \! + [-]$~\cite{Lebed:2017min}.
The current observed properties of the $J^{PC} \! = \! 1^{--}$ ($Y$)
states identified with the multiplet $\Sigma^+_g(1P)$, whose
spectroscopy is analyzed extensively in Ref.~\cite{Giron:2020fvd},
are summarized in Table~\ref{tab:Expt}.
\begin{table*}
  \caption{$J^{PC} \! = \! 1^{--}$ charmoniumlike exotic-meson
  candidates catalogued by the Particle Data Group
  (PDG)~\cite{Zyla:2020zbs}, which are identified with specific
  states within the $\Sigma^+_g(1P)$ multiplet of the dynamical
  diquark model, as summarized by the cases presented in
  Ref.~\cite{Giron:2020fvd} and repeated in
  Sec.~\ref{sec:pwave_review}.  Both the particle name most commonly
  used in the literature and its label as given in the PDG are
  listed.}
\label{tab:Expt}
\centering
\setlength{\extrarowheight}{1.2ex}
\begin{tabular}{cccccc}
\hline\hline
    Particle
        & PDG label
            & $I^{G}J^{PC}$
                & Mass [MeV]
                    & Width [MeV]
                        & Production and decay \\ \hline
    $Y(4220)$ & $\psi(4230)$
        & $0^-1^{--}$
            & $4218^{+5}_{-4}$
                & $59^{+12}_{-10}$
                    & $e^+e^- \to Y$; $Y \to \left\{ \begin{array}{l}
                     \omega \chi_{c0} \\
                     \eta J/\psi\\
                     \pi^+ \pi^- h_c \\
                     \pi^+ \pi^- \psi(2S) \\
		     \pi^+ D^0 D^{*-} \\
		     \pi^0 Z_c^0(3900) \\
		     \gamma X(3872)\\
		\end{array} \right. $ \\ %
    $Y(4260)$ & $\psi(4260)$
        & $0^-1^{--}$
            & $4230 \pm 8$
                & $55 \pm 19$
                    & $e^+e^- \to \gamma Y$ or $Y$; $Y \to \left\{ \begin{array}{l}
                        \pi^+ \pi^- J/\psi \\
                        f_0(980) J/\psi \\
                       \pi^\mp Z_c^\pm (3900) \\
                       K^+ K^- J/\psi \\
                       \gamma X(3872) \end{array} \right. $ \\ %
    $Y(4360)$ & $\psi(4360)$
        & $0^- 1^{--}$
            & $4368 \pm 13$
                & $96 \pm 7$
                    & $e^+e^- \to \gamma Y$ or $Y$; $Y \to \bigg\{ \begin{array}{l}
                        \pi^+ \pi^- \psi(2S) \\
                        \pi^0 \pi^0 \psi(2S) \end{array}$ \\ %
    $Y(4390)$ & $\psi(4390)$
        & $0^- 1^{--}$
            & $4392 \pm 7$
                & $140^{+16}_{-21}$
                    & $e^+e^- \to Y$; $Y \to \bigg\{ \begin{array}{l}
			\eta J/\psi \\		        
			\pi^+ \pi^- h_c \end{array}$ \\ %
    $Y(4660)$ & $\psi(4660)$
        & $0^- 1^{--}$
            & $4643 \pm 9$
                & $72 \pm 11$
                    & $e^+e^- \to \left\{ \begin{array}{l}
                        \gamma Y; Y \to \pi^+ \pi^- \psi(2S) \\
                        \ \; Y; Y \to \Lambda^+_c \Lambda^-_c
                        \end{array} \right.$ \\ \hline\hline
\end{tabular}
\end{table*}

\section{Theoretical Review of $P$-Wave Exotic States}
\label{sec:pwave_review}

The full spectroscopy of diquark-antidiquark ($\de$-$\bde$)
tet\-raquarks and diquark-antitriquark ($\de$-$\bt$) pentaquarks
connected by a gluonic field of arbitrary excitation quantum numbers,
and including arbitrary orbital excitations between the $\de$-$\bde$
or $\de$-$\bt$ pair, is presented in Ref.~\cite{Lebed:2017min}.
As discussed in that work, the gluonic-field excitations combined
with the quasiparticle sources $\de,\bde,\bt$ produce states
analogous to ordinary quarkonium hybrids; therefore, these states
may likewise be classified according to the quantum numbers provided
by BO-approximation static gluonic-field potentials.  The numerical
studies of Ref.~\cite{Giron:2019bcs} show that the exotic analogues
to hybrid quarkonium states lie above the exotic states within the
corresponding BO ground-state potential $\Sigma^+_g$ by at least
1~GeV (just as for conventional quarkonium).  Since the entire range
of observed hidden-charm exotic candidates [not counting $c\bar c
c\bar c$ candidates such as $X(6900)$] spans only about
800~MeV~\cite{Lebed:2016hpi}, it is very likely that all known
hidden-charm exotic states occupy energy levels within the
$\Sigma^+_g$ BO potential.  All known $\ccqq$ candidates can be
accommodated by the lowest $\Sigma^+_g$ levels: $1S$, $1P$, $2S$,
$1D$, and $2P$, in order of increasing mass~\cite{Giron:2019bcs}.

A detailed enumeration of the possible $\QQ q\bar q^\prime$ states,
in which the light quarks $q$,$\bar q^\prime$ do not necessarily
carry the same flavor, is straightforward for the $S$ wave.  Assuming
zero relative orbital angular momenta between the quarks, any two
naming conventions for the states differ only by the order in which
the 4 quark spins are coupled.  In the diquark basis, defined by
coupling in the order $(qQ) \! + \! (\bar q \bar Q)$, the 6 possible
states are denoted by~\cite{Maiani:2014aja}:
\begin{eqnarray}
J^{PC} = 0^{++}: & \ & X_0 \equiv \left| 0_\de , 0_\bde \right>_0 \,
, \ \ X_0^\prime \equiv \left| 1_\de , 1_\bde \right>_0 \, ,
\nonumber \\
J^{PC} = 1^{++}: & \ & X_1 \equiv \frac{1}{\sqrt 2} \left( \left|
1_\de , 0_\bde \right>_1 \! + \left| 0_\de , 1_\bde \right>_1 \right)
\, ,
\nonumber \\
J^{PC} = 1^{+-}: & \ & \, Z \  \equiv \frac{1}{\sqrt 2} \left( \left|
1_\de , 0_\bde \right>_1 \! - \left| 0_\de , 1_\bde \right>_1 \right)
\, ,
\nonumber \\
& \ & \, Z^\prime \equiv \left| 1_\de , 1_\bde \right>_1 \, ,
\nonumber \\
J^{PC} = 2^{++}: & \ & X_2 \equiv \left| 1_\de , 1_\bde \right>_2 \,
,
\label{eq:Swavediquark}
\end{eqnarray}
where outer subscripts indicate total quark spin $S$.  The same
states may be expressed in any other basis by using angular momentum
recoupling coefficients in the form of the relevant $9j$ symbol.  For
the purposes of this work, the most useful alternate basis is that of
definite heavy-quark (and light-quark) spin, $(\QQ) \! + \! (\qq)$:
\begin{eqnarray}
\lefteqn{\left< (s_q \, s_{\bar q}) s_\qq , (s_Q \, s_{\bar Q}) s_\QQ
, S \, \right| \left. (s_q \, s_Q) s_\de , (s_{\bar q} \, s_{\bar Q})
s_\bde , S \right> } & & \nonumber \\
& = & \left( [s_\qq] [s_\QQ] [s_\de] [s_\bde] \right)^{1/2}
\left\{ \begin{array}{ccc} s_q & s_{\bar q} & s_\qq \\
s_Q & s_{\bar Q} & s_\QQ \\ s_\de & s_\bde & S \end{array} \! \right\}
\, , \ \ \label{eq:9jTetra}
\end{eqnarray}
where $[s] \! \equiv \! 2s + 1$ denotes the multiplicity of a
spin-$s$ state.  Using Eqs.~(\ref{eq:Swavediquark}) and
(\ref{eq:9jTetra}), one then obtains
\begin{eqnarray}
J^{PC} = 0^{++}: & \ & X_0 = \frac{1}{2} \left| 0_\qq , 0_\QQ
\right>_0 + \frac{\sqrt{3}}{2} \left| 1_\qq , 1_\QQ \right>_0 \, ,
\nonumber \\
& & X_0^\prime = \frac{\sqrt{3}}{2} \left| 0_\qq , 0_\QQ
\right>_0 - \frac{1}{2} \left| 1_\qq , 1_\QQ \right>_0 \, , 
\nonumber \\
J^{PC} = 1^{++}: & \ & X_1 = \left| 1_\qq , 1_\QQ \right>_1 \, ,
\nonumber \\
J^{PC} = 1^{+-}: & \ & \, Z \; = \frac{1}{\sqrt 2} \left( \left| 
1_\qq , 0_\QQ \right>_1 \! - \left| 0_\qq , 1_\QQ \right>_1 \right)
\, , \nonumber \\
& \ & \, Z^\prime = \frac{1}{\sqrt 2} \left( \left| 1_\qq ,
0_\QQ \right>_1 \! + \left| 0_\qq , 1_\QQ \right>_1 \right) \, ,
\nonumber \\
J^{PC} = 2^{++}: & \ & X_2 = \left| 1_\qq , 1_\QQ \right>_2 \, .
\label{eq:SwaveQQ}
\end{eqnarray}
Once light-quark flavor is included, one obtains 12 states: 6 each
with $I \! = \! 0$ and $I \! = \! 1$, and spin structures in the form
of Eqs.~(\ref{eq:Swavediquark}) or (\ref{eq:SwaveQQ}).\footnote{If
strange quarks are included, one obtains 6 SU(3)$_{\rm flavor}$
octets and 6 singlets.}  Using these states and the most minimal
$3$-parameter Hamiltonian [the $M_0$, $\kappa_{qQ}$, and $V_0$ terms
of Eq.~(\ref{eq:FullHam}) below],
Refs.~\cite{Giron:2019cfc,Giron:2020qpb} calculate the masses of all
$12$ $S$-wave states in the hidden-charm and hidden-bottom sectors
using known masses of $X(3872)$, $Z_c(3900)$, and $Z_c(4020)$ for
the former; and the known masses of $Z_b(10610)$, $Z_b(10650)$, and
their relative $h_b$ to $\Upsilon$ branching fractions for the
latter.  These results incorporate isospin dependence (the $V_0$
term), a feature not explicitly integrated into other diquark models.

Reference~\cite{Giron:2020fvd} extends this analysis by examining the
$P$-wave multiplet, whose mass spectrum is dictated by the most
minimal $5$-parameter Hamiltonian:
\begin{eqnarray}
H & = & M_0 + 2 \kappa_{qQ} ({\bf s}_q \! \cdot \! {\bf s}_Q +
{\bf s}_{\bar q} \! \cdot \! {\bf s}_{\bar Q}) + V_{LS} \,
{\bf L} \cdot {\bf S} \nonumber \\ & & + V_0 \, {\bm \tau}_q
\! \cdot \! {\bm \tau}_{\bar q} \; {\bm \sigma}_q \! \cdot \!
{\bm \sigma}_{\bar q} + V_T \, {\bm \tau}_q \! \cdot
\! {\bm \tau}_{\bar q} \; S_{12}^{(\qq)} \, ,
\label{eq:FullHam}
\end{eqnarray}
where $M_0$ is the common mass of the multiplet, $\kappa_{qQ}$
represents the strength of the spin-spin coupling within each
diquark, $V_{LS}$ is the spin-orbit coupling strength, $V_0$ is the
isospin-dependent coupling,\footnote{$V_0$ in Eq. (\ref{eq:FullHam})
is analogous to the axial coupling in $NN\pi$ interactions.} $V_T$
represents the tensor coupling, and $S_{12}^{(\qq)}$  is the tensor
operator defined as 
\begin{equation}
\label{eq:Tensor}
S_{12}^{(\qq)} \equiv 3 \, {\bm \sigma}_q \! \cdot {\bm r} \,
{\bm \sigma}_{\bar{q}} \! \cdot {\bm r} / r^2 - {\bm \sigma}_q \!
\cdot {\bm \sigma}_{\bar{q}} \, .
\end{equation}
The well-known tabulated expressions for matrix elements of
$S_{12}^{(\qq)}$~({\it e.g.}, in Ref.~\cite{Shalit:1963nuclear})
directly apply neither in the basis of $s_\qq, s_\QQ$ spins nor
$s_\de, s_\bde$ spins, but rather refer to the basis of total
light-quark angular momentum $J_\qq$:
\begin{equation} \label{eq:JqqDef}
{\bf J}_\qq \equiv {\bf L}_\qq + {\bf s}_\qq \, .
\end{equation}
Assuming that $\de$ and $\bde$ have no internal orbital excitation
so that $L_\qq \! = \! L$, the matrix elements of $S_{12}^{(\qq)}$
are most easily computed in the $J_\qq$ basis, with results that are
then related back to the $s_\qq, s_\QQ$ basis by means of recoupling
using $6j$ symbols:
\begin{eqnarray}
\lefteqn{\mathcal{M}_{J_\qq} \equiv
\left< (L,s_\qq),J_\qq,s_\QQ, J | L,(s_{\qq},s_{\QQ}), S, J \right>
} & & \nonumber\\
= & & (-1)^{L+s_\qq+s_\QQ+J}\sqrt{[J_\qq][S]}
\left\{ \begin{array}{ccc}
L & s_\qq & J_\qq\\
s_\QQ & J & S 
\end{array} \right\} . \label{eq:JqqCoef}
\end{eqnarray}
Using this expression, $S_{12}^{(\qq)}$ matrix elements for all
relevant states are tabulated in Ref.~\cite{Giron:2020fvd}.

The experimental status of the $P$-wave $J^{PC} \! = \! 1^{--}$
exotic candidates remains in flux, with BESIII providing the majority
of the most recent data.  With reference to the information presented
in Table~\ref{tab:Expt}, we have already noted that the analysis of
the BESIII Collaboration~\cite{Ablikim:2016qzw} favors the
interpretation of $Y(4260)$ as a superposition of states, the lowest
component of which is $Y(4220)$.  They identify the higher component
with $Y(4360)$, although the previous mass measurements of this state
given in Table~\ref{tab:Expt} are rather higher, and one of several
scenarios considered in Ref.~\cite{Giron:2020fvd} proposes that
$Y(4360)$ and $Y(4390)$ are the same state, while the higher-mass
component in Ref.~\cite{Ablikim:2016qzw} can be interpreted as a
distinct ``$Y(4320)$''.  Alternately, if the only lower states are
$Y(4220)$, $Y(4360)$, and $Y(4390)$, then $Y(4660)$ becomes the
fourth $I \! = \! 0$, $1^{--}$ candidate state in $\Sigma^+_g(1P)$. 

With the mass spectrum of these charmoniumlike states not yet
entirely settled, Ref.~\cite{Giron:2020fvd} also employs information
on their preferred charmonium decay modes as classified by
heavy-quark spin: $\psi$ ($s_{\QQ} \! = \! 1$) or $h_c$ ($s_{\QQ} \!
= \! 0$).  Assuming heavy-quark spin symmetry as expressed by the
conservation of $s_{\QQ}$ in the decays, the heavy-quark spin content
$P_{s_{\QQ}}$ of each state becomes an invaluable diagnostic in
disentangling the $J^{PC} \! = \! 1^{--}$ spectrum.  For example,
from Table~\ref{tab:Expt} one sees that $Y(4220)$ decays to both
$\psi$ states and $h_c$, while if $Y(4360)$ and $Y(4390)$ are in fact
one state, the same can be said for them as well.
Reference~\cite{Giron:2020fvd} also introduces a parameter $\epsilon$
designed to enforce the goodness-of-fit to a particular value $f$ of
$P_{s_{\QQ}}$, which in the case of $s_{\QQ} \! = \! 0$ reads
\begin{equation}
\Delta \chi^2 = \left( \frac{\ln P_{s_{\QQ} = 0} - \ln f}{\epsilon}
\right)^2 .
\end{equation}
In terms of the parameters $P_{s_{\QQ}}$, $f$, and $\epsilon$, the 5
cases discussed in Ref.~\cite{Giron:2020fvd} designed to represent a
variety of interpretations of the current data are:
\begin{enumerate}

\item $Y(4220)$, $Y(4260)$, $Y(4360)$, $Y(4390)$ masses are as given in
the PDG (Table~\ref{tab:Expt}).  No constraint is placed upon
$P_{s_{\QQ} = 0}^{Y(4220)}$ or $P_{s_{\QQ} =
0}^{Y(4390)}$. \label{JunkFit19}

\item $Y(4220)$, $Y(4260)$, $Y(4360)$, $Y(4390)$ masses are as given in
the \@.  $P_{s_{\QQ} = 0}^{Y(4220)}$ is fit to $f \! = \! \frac 1
3$ with $\epsilon \! = \! 0.1$, and $P_{s_{\QQ} = 0}^{Y(4390)}$ is
unconstrained. \label{JunkFit7B}

\item $Y(4220)$, $Y(4360)$, and $Y(4390)$ masses are as given in
the PDG, while $m_{Y(4260)} \! = \! 4251 \! \pm \! 6$~MeV, which is
the weighted average of the 3 PDG values not including the low BESIII
value~\cite{Ablikim:2016qzw}.  $P_{s_{\QQ} = 0}^{Y(4220)}$ is fit to
$f \! = \! \frac 1 3$ with $\epsilon \! = \! 0.2$, and $P_{s_{\QQ} =
0}^{Y(4390)}$ is fit to $f \! = \! \frac 2 3$ with $\epsilon \! = \!
0.05$. \label{JunkFit9B}

\item $Y(4360)$, $Y(4390)$, and $Y(4660)$ masses are as
given in the PDG, but $Y(4260)$ is assumed not to exist, and
$m_{Y(4220)} \!  = \! 4220.1 \! \pm 2.9$~MeV is the weighted average
of the PDG values combined with the newer BESIII
measurements~\cite{Ablikim:2018vxx,Ablikim:2019apl}.  $P_{s_{\QQ} =
0}$ values are as given in Case~\ref{JunkFit9B}. \label{JunkFit21B}

\item $m_{Y(4220)}$ is as given in Case~\ref{JunkFit21B};
$m_{Y(4260)}$ is as given in Case~\ref{JunkFit9B};
$m_{\text{``$Y(4320)$''}} \! = \! 4320 \! \pm \! 13$~MeV is the lower
BESIII $Y(4360)$ mass measurement from~\cite{Ablikim:2016qzw};
$m_{Y(4390)} \! = \! 4386 \! \pm \! 4$~MeV is the weighted average of
the PDG value and the upper BESIII $Y(4360)$ mass measurement
from~\cite{Ablikim:2017oaf}.   $P_{s_{\QQ} = 0}$ values are as given
in Case~\ref{JunkFit9B}.
\label{JunkFit18B}
\end{enumerate}
  
We previously suggested the importance of heavy-quark spin-symmetry
($s_\QQ$) conservation in the decays of exotics, particularly for
$Z_c(3900)$ and $Z_c(4020)$, but also for several other exotic
candidates that to date have only been observed to decay to
charmonium states carrying one specific value of $s_\QQ$ ({\it e.g.},
to $\psi$ {\em or\/} to $h_c$).  We assume that a state like
$Y(4220)$ is able to decay to channels with either value of $s_\QQ$
due to the initial state being a mixture of $s_\QQ$ eigenstates,
rather than to the value of $s_\QQ$ changing in the decay process
through a heavy-quark spin-symmetry violation.  In addition, in this
analysis we take the well-known radiative transition selection rules
to apply to the light degrees of freedom, which carry the total
angular momentum $J_\qq$ defined in Eq.~(\ref{eq:JqqDef}).  As usual,
the operators defining E1 and M1 transitions transform as $J^{P} \! =
\! 1^-$ and $J^{P} \! = \! 1^+$, respectively.

Explicit expressions for radiative transitions between quarkonium
states (themselves transcribed from textbook atomic-physics formulae)
appear in the literature ({\it e.g.}, Ref.~\cite{Barnes:2005pb}), and
may readily be adapted to the present case.  In particular, the
quarkonium orbital angular momentum $L$ is replaced with $J_\qq$, and
the heavy quark mass $m_Q$ is replaced with the diquark mass $m_\de$.
For E1 partial widths, one has
\begin{widetext}
\begin{equation}\label{eqn:e1width}
\Gamma_{\rm E1}\left(n^{\, 2s_\QQ+1} \!
\left(J^{\vphantom\prime}_{\qq}\right)^{\vphantom\dagger}_J
\rightarrow n'^{\; 2s_\QQ'+1} \!
\left(J'_{\qq} \right)^{\vphantom\dagger}_{J'} +
\gamma\right)=\frac{4}{3}C_{fi} \,
\delta_{s^{\vphantom\prime}_\QQ s_\QQ'} \alpha \, Q_{\de}^2
\left\vert\left\langle \psi_f \middle| r \middle| \psi_i
\right\rangle\right\vert^2E_\gamma^3
\frac{E_f^{(\QQ q\bar{q}')}}{M_i^{(\QQ q\bar{q}')}},
\end{equation}
\end{widetext}
where
\begin{equation}
\label{eq:Cfactors}
C_{fi} \equiv \mathrm{max}\left(J_\qq^{\vphantom\prime},
J^\prime_\qq\right) \left(2J'+1\right)
\begin{Bmatrix}
J^\prime_\qq & J' & s_\QQ\\
J & J_\qq & 1
\end{Bmatrix}^2.
\end{equation}
The labels $i$ and $f$ refer to initial and final states,
respectively.  The initial exotic state $\QQ q\bar q^\prime$, of mass
$M_i^{(\QQ q\bar q^\prime)}$, decays in its rest frame into a final
exotic state with the same flavor content and energy $E_f^{(\QQ q\bar
q^\prime)}$, and a photon of energy $E_\gamma$.  $\alpha$ is the
fine-structure constant.  $\psi$  denotes radial wave functions of
the exotic hadrons, and $r$ is the spatial separation between the
$\de$-$\bde$ pair centers.  $Q_\de$ is the total electric charge (in
units of proton charge) to which the photon couples; in
Ref.~\cite{Chen:2015dig}, the diquarks are treated as pointlike, in
which case one simply takes $Q_\de \! = \! Q_Q \! + \! Q_q$.
Alternately, one may argue that the diquarks $\de$ are of sufficient
spatial extent that the photon couplings to the distinct quarks in $\de$
should add through incoherent diagrams, in which case one takes
$Q_{\de}^2 \! = \! Q_Q^2 \! + \! Q_q^2$.  In our calculation we use
the first option, but note in addition that a $Y$ state, being an
isosinglet, contains an equal superposition of $u$ and $d$ quarks.
We thus take
\begin{equation}
\label{eq:charge}
Q_{\de}^2 \to \frac{1}{2} \left[ \left( Q_c + Q_u \right)^2 +
\left( Q_c + Q_d \right)^2 \right] = \frac{17}{18} \, .
\end{equation}
Other schemes give rise to coefficients that differ from this value
only at $O(1)$.  Corrections that arise from treating the distinct
quarks within each diquark as separated entities, for example through
electromagnetic form factors of the $\de,\bde$ composite
quasiparticles, would be incorporated in this model through the
factor $Q_{\de}^2$.

The corresponding expression for M1 partial widths, involving no
change in parity but a flip of the heavy-quark spin $s_{\QQ}$ (hence
breaking heavy-quark spin symmetry), reads
\begin{widetext}
\begin{equation}\label{eqn:m1width}
\Gamma_{\rm M1}\left(n^{\, 2s_\QQ+1}
\left(J^{\vphantom\prime}_{\qq}\right)^{\vphantom\dagger}_J
\rightarrow n'^{\; 2s_\QQ'+1}
\left(J'_{\qq} \right)^{\vphantom\prime}_{J'} +
\gamma\right)=\frac{4}{3} \frac{2J' +1}{2J_\qq +1}
\delta_{J^{\vphantom\prime}_\qq J'_\qq}
\delta_{s^{\vphantom\prime}_\QQ, \, s_\QQ'\pm 1} \, Q_{\de}^2
\frac{\alpha}{m_{\de}^2}\left\vert\left\langle \psi_f \middle| \psi_i
\right\rangle\right\vert^2E_\gamma^3
\frac{E_f^{(\QQ q\bar{q}')}}{M_i^{(\QQ q\bar{q}')}} .
\end{equation}
\end{widetext}
This expression is presented here for completeness, in light of the
current lack of experimental evidence for such transitions.  However,
in Sec.~\ref{sec:analysis} we use it to calculate the expected
radiative width for the yet-unobserved~\cite{Ablikim:2021rba}
transition $Z_c (4020)^0 \! \to \! \gamma X(3872)$.

As noted above, Eqs.~(\ref{eqn:e1width}) and (\ref{eqn:m1width}) are
almost identical to textbook nonrelativistic results.  The only
exception in each case is the inclusion of a factor $E_f/M_i$ to
represent relativistic phase space associated with recoil of the
final-state hadron.  In fact, Ref.~\cite{Brambilla:2004wf}
discusses several distinct relativistic corrections that could be
included in a more complete study.  Since this work represents the
first attempt to calculate the radiative widths for a spectrum of
states whose experimental interpretation remains ambiguous, we
include only a minimal set of physical effects in the analysis.

Lastly, corrections to the long-wavelength approximation discussed in
Sec.~\ref{sec:pwave_review} that are derived by retaining the full
photon plane-wave factor $\exp ( i \bm{k} \cdot \bm{r} )$ have also
been computed ({\it e.g.}, Ref.~\cite{Brambilla:2004wf}).
Explicitly,  Eqs.~(\ref{eqn:e1width}) and (\ref{eqn:m1width}) are
modified through the substitutions

\begin{equation}
\langle \psi_f | r | \psi_i \rangle \to \left< \psi_f  \! \left|
\frac{3}{k} \left[ \frac{kr}{2} j_0 \left( \frac{kr}{2} \right) - j_1
\left( \frac{kr}{2} \right) \right] \right| \psi_i \right> \, ,
\end{equation}
and
\begin{equation}
\langle \psi_f | \psi_i \rangle \to \left< \psi_f  \! \left| \, j_0
\left( \frac{kr}{2} \right) \right| \psi_i \right> \, ,
\end{equation}
respectively,where $j_0$ and $j_1$ are spherical Bessel 
functions.  The corresponding series expansions of these functions
read
\begin{equation}
\label{eqn:E1corr}
r - \frac{1}{20} k^2 r^3 + O(k^4 r^5) \, ,
\end{equation}
and
\begin{equation}
\label{eqn:M1corr}
1 - \frac{1}{24} k^2 r^2 + O(k^4 r^4) \, .
\end{equation}
Note especially the small subleading-term numerical coefficient in
each case, suggesting that the long-wavelength approximation holds
relatively well even for substantial values of $kr$.  We examine
specific examples in Sec.~\ref{sec:analysis}.

\section{Analysis and Results}\label{sec:analysis}

Possible assignments of observed $Y$ states to members of the
$\Sigma^+_g(1P)$ multiplet in this model are described by the 5 cases
discussed extensively in Ref.~\cite{Giron:2020fvd} and summarized 
in Sec.~\ref{sec:pwave_review}. Of these cases, all have excellent
goodness-of-fit values $\chi^2_{\rm min}/{\rm d.o.f.}$ except Case~3;
however, we argue this case and Case~5 to be the most
phenomenologically relevant ones, since they enforce the important
physical constraint that both $Y(4220)$ and $Y(4390)$ are observed
(see Table~\ref{tab:Expt}) to have substantial couplings to $h_c$
($s_{\QQ} \! = \! 0$).  Since $\Sigma^+_g(1S)$ contains only one
$I \! = \! 0$, $J^{PC} \! = \! 1^{--}$ state with $s_{\QQ} \! = \!
0$, the requirement of providing a substantial component of this
state to both of the well-separated $Y(4220)$ and $Y(4390)$ mass
eigenstates is one of the primary obstacles to achieving a good fit.

Case~5 relieves the tension of Case~3 by identifying, as discussed in
Sec.~\ref{sec:Expt}, a new state ``$Y(4320)$'' from the data of
Ref.~\cite{Ablikim:2016qzw}.  In addition, Cases~1, 2, 3, and 5 all
predict the sole $I \! = \! 1$, $J^{PC} \! = \! 0^{--}$ state in
$\Sigma^+_g(1P)$ to lie in the range 4220--4235~MeV, which agrees
well with the unconfirmed state $Z_c(4240)$ carrying these quantum
numbers that is observed in the LHCb paper~\cite{Aaij:2014jqa}
confirming the existence of $Z_c(4430)$.

Case~4 also satisfies the $Y(4220)/Y(4390)$ $s_{\QQ} \! = \! 0$
criterion, but additionally assigns the rather high-mass $Y(4660)$ to
the $\Sigma^+_g(1P)$ multiplet; the cost is a much higher prediction
($\approx \! 4440$~MeV) for the mass of the $\Sigma^+_g(1P)$ $I \! =
\! 1$, $J^{PC} \! = \! 0^{--}$ state, in conflict with the value of
$m_{Z_c(4240)}$.

We therefore single out the fits of Cases~3 and 5 for the
decomposition of $Y$ states with respect to the total light-quark
angular momentum $J_{\qq}$ in Tables~\ref{tab:case3} and
\ref{tab:case5}, respectively.  For completeness, we also provide the
corresponding information for Cases~1, 2, and 4 in
Table~\ref{tab:cases124}.

\begin{table}
\caption{Decomposition of $Y$ ($I \! = \! 0$, $J^{PC} \! = \!
1^{--}$) charmoniumlike exotic candidates into a basis of good
light-quark spin $s_{\qq}$, heavy-quark spin $s_{\QQ}$, and total
light-quark angular momentum $J_{\qq}$, performed for the 4
experimentally observed candidate states as described in Case~3 above
and in Ref.~\cite{Giron:2020fvd}.  A minus sign on the probability
$(-|P|)$ means that the corresponding amplitude is $-|P|^{1/2}$, the
same convention as is used for Clebsch-Gordan coefficients by the
PDG~\cite{Zyla:2020zbs}.}
\label{tab:case3}
%\centering
\setlength{\extrarowheight}{0.5ex}
\begin{tabular*}{\columnwidth}{@{\extracolsep{\fill}}c c c c c}
\hline\hline
 %\multicolumn{5}{c}{Case 3}\\
 %\hline
Particle &$s_{\qq}$ & $s_{\cc}$ &\;\;$J_{\qq}$\;\;& Probability\\
\hline
$Y(4220)$ & $0$ & $0$ & $0$ & $+0.231$\\
$ $& $1$ & $1$ & $0$ & $+0.012$\\
$ $&  & $ $ & $1$ & $-0.577$\\
$ $&  & $ $ & $2$ & $+0.181$\\
%\hline
$Y(4260)$ & $0$ & $0$ & $0$ &  $+0.061$\\
$ $& $1$ & $1$ & $0$ &  $+0.004$\\
$ $&  & $ $ & $1$ & $+0.352$\\
$ $&  & $ $ & $2$ & $+0.583$\\
%\hline
$Y(4360)$ & $0$ & $0$ & $0$  & $+0.069$\\
$ $& $1$ & $1$ & $0$ & $+0.835$\\
$ $&  & $ $ & $1$ &  $+0.020$\\
$ $&  & $ $ & $2$ &  $-0.075$\\
%\hline
$Y(4390)$ & $0$ & $0$ & $0$ & $+0.638$\\
$ $& $1$ & $1$ & $0$  & $-0.149$\\
$ $&  & $ $ & $1$ & $+0.051$\\
$ $&  & $ $ & $2$ & $-0.161$\\
\hline\hline
\end{tabular*}
\end{table}

\begin{table}
\caption{Decomposition of $Y$ ($I \! = \! 0$, $J^{PC} \! = \!
1^{--}$) charmoniumlike exotic candidates as in Table~\ref{tab:case3},
except now performed for the 4 experimentally observed candidate
states as described in Case~5 above and in
Ref.~\cite{Giron:2020fvd}.}
\label{tab:case5}
%\centering
\setlength{\extrarowheight}{0.5ex}
\begin{tabular*}{\columnwidth}{@{\extracolsep{\fill}}c c c c c}
\hline\hline
%\multicolumn{5}{c}{Case 5}\\
Particle & $s_{\qq}$ & $s_{\cc}$ &\;\;$J_{\qq}$\;\; & Probability\\
\hline
$Y(4220)$ & $0$ & $0$ & $0$ & $-0.264$\\
$ $& $1$ & $1$ & $0$ & $-0.007$\\
$ $&  & $ $ & $1$ & $+0.543$\\
$ $&  & $ $ & $2$ & $-0.186$\\
%\hline
$Y(4260)$ & $0$ & $0$ & $0$ & $+0.060$\\
$ $& $1$ & $1$ & $0$ & $+0.036$\\
$ $&  & $ $ & $1$ & $+0.380$\\
$ $&  & $ $ & $2$ & $+0.523$\\
%\hline
``$Y(4320)$'' & $0$ & $0$ & $0$  & $+0.025$\\
$ $& $1$ & $1$ & $0$  & $+0.870$\\
$ $&  & $ $ & $1$ & $+8 \times 10^{-4}$\\
$ $&  & $ $ & $2$ & $-0.105$\\
%\hline
$Y(4390)$ & $0$ & $0$ & $0$ & $-0.651$\\
$ $& $1$ & $1$ & $0$  & $+0.086$\\
$ $&  & $ $ & $1$  & $-0.076$\\
$ $&  & $ $ & $2$  & $+0.187$\\
\hline
\end{tabular*}
\end{table}

\begin{table*}
\caption{Decomposition of $Y$ ($I \! = \! 0$, $J^{PC} \! = \!
1^{--}$) charmoniumlike exotic candidates as in
Tables~\ref{tab:case3}--\ref{tab:case5}, except now performed for the
4 experimentally observed candidate states as described in Cases~1,
2, and 4 above and in Ref.~\cite{Giron:2020fvd}.}
\label{tab:cases124}
%\centering
\setlength{\extrarowheight}{0.5ex}
\begin{tabular*}{\textwidth}{@{\extracolsep{\fill}}c c c c c | c c c c c | c c c c c}
%\begin{tabular}{c c c c c | @{}>{\hspace*{1ex}} c c c c c | @{}>{\hspace*{1ex}} c c c c c}
\hline\hline
\multicolumn{5}{c|}{Case 1}& \multicolumn{5}{c|}{Case 2}& \multicolumn{5}{c}{Case 4}\\
\hline
Particle & $s_{\qq}$ & $s_{\cc}$ &\;\;$J_{\qq}$\;\; & Probability& Particle & $s_{\qq}$ & $s_{\cc}$ &\;\;$J_{\qq}$\;\; & Probability &Particle & $s_{\qq}$ & $s_{\cc}$ &\;\;$J_{\qq}$\;\; & Probability\\
\hline
$Y(4220)$ & $0$ & $0$ & $0$ & $+0.771$ & $Y(4220)$ & $0$ & $0$ & $0$ & $-0.336$ & $Y(4220)$ & $0$ & $0$ & $0$& $-0.233$\\
$ $& $1$ & $1$ & $0$ & $-0.019$ & $ $ & $1$ & $1$ & $0$ & $+0.032$& $ $& $1$ & $1$ & $0$& $-0.048$\\
$ $&  & $ $ & $1$ & $-0.211$ & $ $ &  & $ $ & $1$ & $+0.631$ & $ $&  & $ $ & $1$& $+0.376$\\
$ $&  & $ $ & $2$ & $-4\times 10^{-7}$ & $ $&  & $ $ & $2$ & $+8\times 10^{-4}$ & $ $&  & $ $ & $2$& $-0.343$\\
%\hline
$Y(4260)$ & $0$ & $0$ & $0$ & $+0.212$ & $Y(4260)$ & $0$ & $0$ & $0$ & $+0.588$ & $Y(4360)$ & $0$ & $0$ & $0$& $-0.119$\\
$ $& $1$ & $1$ & $0$ & $+0.130$ & $ $ & $1$ & $1$ & $0$ & $+0.056$ & $ $& $1$ & $1$ & $0$& $+0.101$\\
$ $&  & $ $ & $1$& $+0.597$ &$ $ &  & $ $ & $1$ & $+0.246$ & $ $&  & $ $ & $1$& $-0.473$\\
$ $ &  & $ $ & $2$ &$+0.062$ & $ $&  & $ $ & $2$ & $+0.109$ &  $ $ &  & $ $ & $2$ & $-0.308$ \\
%\hline
$Y(4360)$ & $0$ & $0$ & $0$ & $+0.006$ & $Y(4360)$ & $0$ & $0$ & $0$ & $-0.046$ & $Y(4390)$ & $0$ & $0$ & $0$& $+0.647$\\
$ $& $1$ & $1$ & $0$ & $+0.252$ & $ $& $1$ & $1$ & $0$ & $-0.117$ & $ $& $1$ & $1$ & $0$ & $+0.003$\\
$ $&  & $ $ & $1$ & $-5\times 10^{-6}$ & $ $&  & $ $ & $1$ & $-0.012$ & $ $&  & $ $ & $1$& $+0.009$\\
$ $&  & $ $ & $2$ & $-0.742$ & $ $&  & $ $ & $2$ & $+0.824$ & $ $&  & $ $ & $2$& $-0.342$\\
%\hline
$Y(4390)$ & $0$ & $0$ & $0$ & $-0.012$ & $Y(4390)$ & $0$ & $0$ & $0$ & $-0.029$ &$Y(4660)$ & $0$ & $0$ & $0$ & $-0.002$\\
$ $& $1$ & $1$ & $0$ & $+0.599$ & $ $ & $1$ & $1$ & $0$ & $+0.795$ & $ $& $1$ & $1$ & $0$ & $+0.848$\\
$ $& $ $ & $ $ & $1$ & $-0.193$ & $ $ &  & $ $ & $1$ & $-0.111$ & $ $&  & $ $ & $1$ & $+0.143$\\
$ $& $ $ & $ $ & $2$ & $+0.196$ & $ $ &  & $ $ & $2$ & $+0.066$ & $ $&  & $ $ & $2$ & $+0.007$\\
\hline\hline
\end{tabular*}
\end{table*}

Using the mass eigenvalues for the $Y$ states in
Table~\ref{tab:Expt}, the state decompositions according to $J_{\qq}$
in Tables~\ref{tab:case3}, \ref{tab:case5}, and \ref{tab:cases124},
the coefficient factors in Eq.~(\ref{eq:Cfactors}), and the effective
squared-charge $Q_\delta^2$ from Eq.~(\ref{eq:charge}), one may
calculate the E1 radiative partial decay widths for $\Sigma^+_g(1P)
\! \to \! \gamma \Sigma^+_g(1S)$ transitions from
Eq.~(\ref{eqn:e1width}).  The only nontrivial new input to the
calculation is that of the transition matrix element $\langle \psi_f
| r | \psi_i \rangle$.  Using the numerical methods for solving
Schr\"{o}dinger equations developed in Ref.~\cite{Giron:2019bcs}, and
particularly the fits performed in Ref.~\cite{Giron:2020qpb} to
obtain the fine structure of the $\Sigma^+_g(1S)$ multiplet, the
optimal diquark mass is found to be
\begin{equation}
\label{eq:diquarkmass}
m_{\de} = m_{\bde} = 1.933 \pm 0.005 \, {\rm GeV} \, ,
\end{equation}
as one varies over the static gluonic-field potentials $\Sigma^+_g$
obtained in the lattice calculations of Refs.~\cite{Juge:1997nc,
Juge:1999ie,Juge:2002br,Morningstar:2019,Capitani:2018rox}.  We then
compute the relevant matrix element to be
\begin{equation}
\label{eq:TransMatEl}
\left< \psi_f (1S) \left| r \right| \psi_i (1P) \right> = 0.402 \pm
0.001 \, {\rm fm} \, .
\end{equation}
Note in particular that this numerical input appears in all
$\Sigma^+_g(1P) \! \to \! \gamma \Sigma^+_g(1S)$ transitions, not
simply those of $Y \! \to \! \gamma X(3872)$ that are compiled
according to the 5 cases in Table~\ref{tab:partialwidth}.  Moreover,
Table~\ref{tab:partialwidth} and additional calculated width values
presented subsequently in this work exhibit only central values for
$\Gamma$; the small uncertainties in Eqs.~(\ref{eq:diquarkmass}) and
(\ref{eq:TransMatEl}) only refer to variation over different lattice
simulations, and do not take into account other much more significant
potential sources of uncertainty, such as effects due to finite
diquark size.  Nevertheless, such effects were
shown~\cite{Giron:2020qpb} to change expectation values like $\langle
r \rangle$ no more than 10\%, a value that we adopt as a benchmark
uncertainty for all $\Gamma$ values computed here.

Also noteworthy is the magnitude of $k \left< \psi_f (1S) \! \left| r
\right| \! \psi_i (1P) \right>$ for each case, which provides an
indication of the reliability of the long-wavelength approximation.
Indeed, for $Y(4220)$, $k \! = \! 334$~MeV, and using
Eq.~(\ref{eq:TransMatEl}) gives $kr \! \to \! 0.680$, while the
corresponding value for $Y(4660)$ ($k \! = \! 699$~MeV) is 1.424.
However, the same simulations as in Eq.~(\ref{eq:TransMatEl}) also
produce
\begin{equation}
\left< \psi_f (1S) \left| r^3 \right| \psi_i (1P) \right> = 0.135 \pm
0.001 \, {\rm fm}^3 \, ,
\end{equation}
from which one computes the relative magnitude of the first
correction term in Eq.~(\ref{eqn:E1corr}) to be only 0.033 for
$Y(4220)$ and, surprisingly, only 0.300 for $Y(4660)$.

\begin{table*}
\caption{Radiative E1 partial widths (in keV) to $\gamma X(3872)$
calculated using Eq.~(\ref{eqn:e1width}), for the 5 cases of possible
$Y$ state assignments defined above and in Ref.~\cite{Giron:2020fvd}.
For each case, note that two of the $Y$ states (indicated by dashes)
are assumed either not to exist or not to belong to the
$\Sigma^+_g(1P)$ multiplet.}
\centering
\setlength{\extrarowheight}{1.2ex}
\begin{tabular}{c | r | r | r | r | r | r }
\hline\hline
Case &$Y(4220)$ & $Y(4260)$ & ``$Y(4320)$'' & $Y(4360)$ & $Y(4390)$ & $Y(4660)$\\
\hline
%Case & \\ %\multicolumn{6}{c|}{$\Gamma_{\rm E1}$ (MeV)}\\
%\hline
$1$ & $30.4$ & $145.6$ & {--- \ } & $721.3$ & $981.8$ & {--- \ } \\
$2$ & $81.1$ & $80.6$ & {--- \ } & $616.0$ & $1127.0$ & {--- \ } \\
$3$ & $105.1$ & $211.2$ & {--- \ } & $1016.1$ & $319.2$ & {--- \ } \\
$4$ & $136.0$ & {--- \ } & {--- \ } & $432.1$ & $231.6$ & $3363.9$ \\
$5$ & $102.4$ & $216.2$ & $807.6$ & {--- \ } & $253.8$ & {--- \ } \\
\hline
\end{tabular}
\label{tab:partialwidth}
\end{table*}

One observes from Table~\ref{tab:partialwidth} that the widths
$\Gamma_{Y(4220) \to \gamma X(3872)}$ and
$\Gamma_{Y(4260) \to \gamma X(3872)}$ assume almost the same
values in Cases~3 and 5 (102--105~keV and 211--216~keV,
respectively). $\Gamma_{Y(4390) \to \gamma X(3872)}$ also exhibits
fairly modest variation, from 254--319~keV\@.  Indeed, some of the
large radiative width values in Table~\ref{tab:partialwidth}, such as
3.4~MeV for $Y(4660) \! \to \! \gamma X(3872)$ in Case 4, can serve
as vital criteria for eliminating possible assignments of $Y$ states
to the $1P$ multiplet: Glancing at the measured total
$\Gamma_{Y(4660)}$ in Table~\ref{tab:Expt}, one sees that were
$Y(4660)$ truly a $1P$ state, then its large phase space for
radiative decay to $X(3872)$ [evident from the $E_\gamma^3$ factor of
Eq.~(\ref{eqn:e1width})] would generate a radiative branching
fraction of at least several percent.

The transition matrix element of Eq.~(\ref{eq:TransMatEl}) has
already been noted to apply to all $\Sigma^+_g(1P) \! \to \! \gamma
\Sigma^+_g(1S)$ transitions.  The only observed hidden-charm
tetraquark candidates with $P \! = \! -$ apart from the $Y$ states
are $Z_c(4240)$ and $Y(4626)$; the latter has thus far been observed
to decay only to various $D_s$ meson pairs~\cite{Jia:2019gfe,
Jia:2020epr}, and therefore is very likely a $c\bar c s\bar s$
state~\cite{Giron:2020qpb}.  As for $Z_c(4240)$, only its charged
isobar has yet been observed, but assuming the existence of a
degenerate $Z_c(4240)^0$, one may input its quantum numbers
$s_{\QQ} \! = \! 1$, $J_{\qq} \! = \! 1$, $J \! = \!
0$~\cite{Giron:2020fvd} into Eq.~(\ref{eqn:e1width}) to obtain
\begin{equation}
\Gamma \left[ Z_c (4240)^0 \to \gamma X(3872) \right] = 503 \
{\rm keV} \, .
\end{equation}

Lastly, we noted with Eq.~(\ref{eqn:m1width}) that M1 transitions
occur only with a flip of the heavy-quark spin.  Such is the case for
the $\Sigma^+_g(1S) \! \to \! \gamma \Sigma^+_g(1S)$ transition
$Z_c(4020)^0 \! \to \! \gamma X(3872)$ ($s_{\QQ} \! = \! 0 \! \to
s_{\QQ} \! = \! 1$).  Using Eq.~(\ref{eq:diquarkmass}), we calculate
\begin{equation}
\Gamma \left[ Z_c (4020)^0 \to \gamma X(3872) \right] = 7.91 \
{\rm keV} \, ,
\end{equation}
noting from Eq.~(\ref{eqn:m1width}) that the underlying matrix
element $\langle \psi_f | \psi_i \rangle \! = \! 1$ since both states
share the same radial wave function.  In comparison, the molecular
model, in which $X(3872)$ and $Z_c(4020)$ are $D^0 \bar D^{*0} \! +
\! {\bar D}^0 D^{*0}$ and $D^* \bar D^*$ bound states, respectively,
and the decay $Z_c(4020)^0 \! \to \! \gamma X(3872)$ proceeds via
$D^{*0} \! \to \! \gamma D^0$, produces a rather larger radiative
width: The calculation of Ref.~\cite{Voloshin:2019ivc} predicts a
branching fraction of about $5 \! \times \! 10^{-3}$, which for
$\Gamma_{Z_c(4020)} \! = \! 13 \! \pm \! 5$~MeV~\cite{Zyla:2020zbs}
amounts to at least 40~keV.

In light of our investigations for $\Sigma^+_g (1P) \! \to \!
\Sigma^+_g(1S)$ E1 transitions, the long-wavelength approximation for
M1 transitions within the single multiplet $\Sigma^+_g(1S)$ is
undoubtedly satisfactory [for example, in $Z_c(4020)^0 \! \to \!
\gamma X(3872)$, $k$ is only 150~MeV].  Indeed, one may press the
approximation of Eq.~(\ref{eqn:M1corr}) to consider a transition that
is forbidden in the long-wavelength limit of Eq.~(\ref{eqn:m1width})
due to the orthogonality of wave functions, such as $\Sigma^+_g(2S)
\! \to \! \gamma \Sigma^+_g(1S)$.  Assuming that $Z_c^0(4430)$ is the
$2S$ partner to $Z_c^0(4020)$, then $Z_c^0(4430) \! \to \! \gamma
X(3872)$ has $k \! \approx \! 565$~MeV, while we compute
\begin{equation}
\left< \psi_f (1S) \left| r^2 \right| \psi_i (2S) \right> = 0.152 \pm
0.001 \, {\rm fm}^2 \, ,
\end{equation}
and the first nontrivial term of Eq.~(\ref{eqn:M1corr}) evaluates to
$-0.052$.  Using this correction in Eq.~(\ref{eqn:m1width}) leads to
a radiative width $\Gamma_\gamma \! \simeq \! 1$~keV, to be compared
with $\Gamma_{Z_c^+(4430)} \! \approx \!
180$~MeV~\cite{Zyla:2020zbs}.  The observation a radiative transition
with such a small branching fraction is not impossible, but likely
will not occur in the near future.

\section{Conclusions}\label{sec:conclusions}

In this paper we have calculated exotic-to-exotic had\-ronic
radiative transitions using the dynamical diquark model.  The most
phenomenologically relevant final state is $X(3872)$, which is a
member of the model's hidden-charm ground-state multiplet
$\Sigma^+_g(1S)$.  We use the results from a recent
study~\cite{Giron:2020fvd} of this model for the lowest $P$-wave
multiplet [$\Sigma^+_g(1P)$] of hidden-charm tetraquark states, in
which the $\Sigma^+_g(1P)$ states are identified with the observed
$I \! = \! 0$, $J^{PC} \! = \! 1^{--}$ ($Y$) states according to a
variety of scenarios, based upon both their mass spectra and
preferred decay modes to eigenstates of heavy-quark spin ({\it e.g.},
$J/\psi$ {\it vs.} $\! h_c$).  We calculate E1 and M1 transition
amplitudes for $\Sigma_g^{+}(1P) \! \to \! \gamma\Sigma_g^{+}(1S)$
and $\Sigma_g^{+}(1S) \! \to \! \gamma \Sigma_g^{+}(1S)$ processes,
respectively, and present corresponding values for the radiative
decay widths of a number of particular exclusive channels.

This analysis shows that if $Y(4220)$ and $X(3872)$ have a similar
underlying diquark structure, then one expects $\Gamma_{Y(4220) \to
\gamma X(3872)} \! \approx \! 100$~keV\@.   Moreover, similar values
(albeit somewhat larger due to increased $\gamma$ phase space) are
expected for the heavier $Y$ states in $\Sigma^+_g(1P)$.  The extreme
possibility of $Y(4660)$ belonging to the $1P$ multiplet would lead
to a $\gamma X(3872)$ branching fraction of several percent, and so
the absence of such a remarkably large signal would appear to
relegate $Y(4660)$ instead to the $\Sigma^+_g(2P)$ multiplet.

Furthermore, we found that the observed but uncon\-firmed
$Z_c(4240)$, a candidate for the sole $I \! = \! 1$, $J^{PC}=0^{--}$
state in $\Sigma^+_g(1P)$, should have a substantial ($\approx \!
500$~keV) radiative decay width to $X(3872)$ through its neutral
isobar, and therefore this decay is a good candidate for future
experimental investigation.  Indeed, many of the $\Sigma^+_g(1P)$
states have not yet been observed, offering multiple potential future
tests of the model.

M1 transitions within a single multiplet, such as $Z_c(4020)^0 \! \to
\! \gamma X(3872)$, produce much narrower widths ($< \! 10$~keV in
this model), and can provide sensitive tests of substructure
({\it e.g.}, diquarks {\it vs.} meson molecules).

One may also study exotic-to-exotic radiative transitions in other
heavy-quark sectors ({\it e.g.}, hidden-bottom or $c\bar c s\bar s$
exotics).  Indeed, Eqs.~(\ref{eqn:e1width}) and (\ref{eqn:m1width})
are general for any tetraquark state in the diquark-antidiquark
configuration. For example, Ref.~\cite{Giron:2020qpb} calculates the
mass of $X_b$ [the hidden-bottom analogue to $X(3872)$] to lie in a
rather narrow range $m_{X_b}\in[10598,10607]$ MeV, only slightly
below the observed $Z_b(10610)^0$.  The M1 transition $Z_b(10610)^0
\! \to \! \gamma X_b$ is thus expected from Eq.~(\ref{eqn:m1width})
to produce a tiny [$O({\rm eV})$ or less] radiative width, owing to
not only the small phase space, but also the larger ($b$-containing)
diquark mass.  We conclude that even very coarse experimental results
in other sectors can be decisive in verifying or falsifying
particular models.

\begin{acknowledgments}
This work was supported by the National Science Foundation (NSF) under 
Grant No.\ PHY-1803912.
\end{acknowledgments}

\bibliographystyle{apsrev4-1}
\bibliography{diquark}

%merlin.mbs apsrev4-1.bst 2010-07-25 4.21a (PWD, AO, DPC) hacked
%Control: key (0)
%Control: author (72) initials jnrlst
%Control: editor formatted (1) identically to author
%Control: production of article title (-1) disabled
%Control: page (0) single
%Control: year (1) truncated
%Control: production of eprint (0) enabled
\begin{thebibliography}{50}%
\makeatletter
\providecommand \@ifxundefined [1]{%
 \@ifx{#1\undefined}
}%
\providecommand \@ifnum [1]{%
 \ifnum #1\expandafter \@firstoftwo
 \else \expandafter \@secondoftwo
 \fi
}%
\providecommand \@ifx [1]{%
 \ifx #1\expandafter \@firstoftwo
 \else \expandafter \@secondoftwo
 \fi
}%
\providecommand \natexlab [1]{#1}%
\providecommand \enquote  [1]{``#1''}%
\providecommand \bibnamefont  [1]{#1}%
\providecommand \bibfnamefont [1]{#1}%
\providecommand \citenamefont [1]{#1}%
\providecommand \href@noop [0]{\@secondoftwo}%
\providecommand \href [0]{\begingroup \@sanitize@url \@href}%
\providecommand \@href[1]{\@@startlink{#1}\@@href}%
\providecommand \@@href[1]{\endgroup#1\@@endlink}%
\providecommand \@sanitize@url [0]{\catcode `\\12\catcode `\$12\catcode
  `\&12\catcode `\#12\catcode `\^12\catcode `\_12\catcode `\%12\relax}%
\providecommand \@@startlink[1]{}%
\providecommand \@@endlink[0]{}%
\providecommand \url  [0]{\begingroup\@sanitize@url \@url }%
\providecommand \@url [1]{\endgroup\@href {#1}{\urlprefix }}%
\providecommand \urlprefix  [0]{URL }%
\providecommand \Eprint [0]{\href }%
\providecommand \doibase [0]{http://dx.doi.org/}%
\providecommand \selectlanguage [0]{\@gobble}%
\providecommand \bibinfo  [0]{\@secondoftwo}%
\providecommand \bibfield  [0]{\@secondoftwo}%
\providecommand \translation [1]{[#1]}%
\providecommand \BibitemOpen [0]{}%
\providecommand \bibitemStop [0]{}%
\providecommand \bibitemNoStop [0]{.\EOS\space}%
\providecommand \EOS [0]{\spacefactor3000\relax}%
\providecommand \BibitemShut  [1]{\csname bibitem#1\endcsname}%
\let\auto@bib@innerbib\@empty
%</preamble>
\bibitem [{\citenamefont {Lebed}\ \emph {et~al.}(2017)\citenamefont {Lebed},
  \citenamefont {Mitchell},\ and\ \citenamefont {Swanson}}]{Lebed:2016hpi}%
  \BibitemOpen
  \bibfield  {author} {\bibinfo {author} {\bibfnamefont {R.}~\bibnamefont
  {Lebed}}, \bibinfo {author} {\bibfnamefont {R.}~\bibnamefont {Mitchell}}, \
  and\ \bibinfo {author} {\bibfnamefont {E.}~\bibnamefont {Swanson}},\ }\href
  {\doibase 10.1016/j.ppnp.2016.11.003} {\bibfield  {journal} {\bibinfo
  {journal} {Prog.\ Part.\ Nucl.\ Phys.}\ }\textbf {\bibinfo {volume} {{\bf
  93}}},\ \bibinfo {pages} {143} (\bibinfo {year} {2017})},\ \Eprint
  {http://arxiv.org/abs/1610.04528} {arXiv:1610.04528 [hep-ph]} \BibitemShut
  {NoStop}%
%%CITATION = ARXIV:1610.04528;%%
\bibitem [{\citenamefont {Chen}\ \emph {et~al.}(2016)\citenamefont {Chen},
  \citenamefont {Chen}, \citenamefont {Liu},\ and\ \citenamefont
  {Zhu}}]{Chen:2016qju}%
  \BibitemOpen
  \bibfield  {author} {\bibinfo {author} {\bibfnamefont {H.-X.}\ \bibnamefont
  {Chen}}, \bibinfo {author} {\bibfnamefont {W.}~\bibnamefont {Chen}}, \bibinfo
  {author} {\bibfnamefont {X.}~\bibnamefont {Liu}}, \ and\ \bibinfo {author}
  {\bibfnamefont {S.-L.}\ \bibnamefont {Zhu}},\ }\href {\doibase
  10.1016/j.physrep.2016.05.004} {\bibfield  {journal} {\bibinfo  {journal}
  {Phys.\ Rep.}\ }\textbf {\bibinfo {volume} {{\bf 639}}},\ \bibinfo {pages}
  {1} (\bibinfo {year} {2016})},\ \Eprint {http://arxiv.org/abs/1601.02092}
  {arXiv:1601.02092 [hep-ph]} \BibitemShut {NoStop}%
%%CITATION = ARXIV:1601.02092;%%
\bibitem [{\citenamefont {Hosaka}\ \emph {et~al.}(2016)\citenamefont {Hosaka},
  \citenamefont {Iijima}, \citenamefont {Miyabayashi}, \citenamefont {Sakai},\
  and\ \citenamefont {Yasui}}]{Hosaka:2016pey}%
  \BibitemOpen
  \bibfield  {author} {\bibinfo {author} {\bibfnamefont {A.}~\bibnamefont
  {Hosaka}}, \bibinfo {author} {\bibfnamefont {T.}~\bibnamefont {Iijima}},
  \bibinfo {author} {\bibfnamefont {K.}~\bibnamefont {Miyabayashi}}, \bibinfo
  {author} {\bibfnamefont {Y.}~\bibnamefont {Sakai}}, \ and\ \bibinfo {author}
  {\bibfnamefont {S.}~\bibnamefont {Yasui}},\ }\href {\doibase
  10.1093/ptep/ptw045} {\bibfield  {journal} {\bibinfo  {journal} {Prog.\
  Theor.\ Exp.\ Phys.}\ }\textbf {\bibinfo {volume} {{\bf 2016}}},\ \bibinfo
  {pages} {062C01} (\bibinfo {year} {2016})},\ \Eprint
  {http://arxiv.org/abs/1603.09229} {arXiv:1603.09229 [hep-ph]} \BibitemShut
  {NoStop}%
%%CITATION = ARXIV:1603.09229;%%
\bibitem [{\citenamefont {Esposito}\ \emph {et~al.}(2017)\citenamefont
  {Esposito}, \citenamefont {Pilloni},\ and\ \citenamefont
  {Polosa}}]{Esposito:2016noz}%
  \BibitemOpen
  \bibfield  {author} {\bibinfo {author} {\bibfnamefont {A.}~\bibnamefont
  {Esposito}}, \bibinfo {author} {\bibfnamefont {A.}~\bibnamefont {Pilloni}}, \
  and\ \bibinfo {author} {\bibfnamefont {A.}~\bibnamefont {Polosa}},\ }\href
  {\doibase https://doi.org/10.1016/j.physrep.2016.11.002} {\bibfield
  {journal} {\bibinfo  {journal} {Phys.\ Rep.}\ }\textbf {\bibinfo {volume}
  {668}},\ \bibinfo {pages} {1} (\bibinfo {year} {2017})},\ \Eprint
  {http://arxiv.org/abs/1611.07920} {arXiv:1611.07920 [hep-ph]} \BibitemShut
  {NoStop}%
\bibitem [{\citenamefont {Guo}\ \emph {et~al.}(2018)\citenamefont {Guo},
  \citenamefont {Hanhart}, \citenamefont {Mei{\ss}ner}, \citenamefont {Wang},
  \citenamefont {Zhao},\ and\ \citenamefont {Zou}}]{Guo:2017jvc}%
  \BibitemOpen
  \bibfield  {author} {\bibinfo {author} {\bibfnamefont {F.-K.}\ \bibnamefont
  {Guo}}, \bibinfo {author} {\bibfnamefont {C.}~\bibnamefont {Hanhart}},
  \bibinfo {author} {\bibfnamefont {U.-G.}\ \bibnamefont {Mei{\ss}ner}},
  \bibinfo {author} {\bibfnamefont {Q.}~\bibnamefont {Wang}}, \bibinfo {author}
  {\bibfnamefont {Q.}~\bibnamefont {Zhao}}, \ and\ \bibinfo {author}
  {\bibfnamefont {B.-S.}\ \bibnamefont {Zou}},\ }\href {\doibase
  10.1103/RevModPhys.90.015004} {\bibfield  {journal} {\bibinfo  {journal}
  {Rev.\ Mod.\ Phys.}\ }\textbf {\bibinfo {volume} {{\bf 90}}},\ \bibinfo
  {pages} {015004} (\bibinfo {year} {2018})},\ \Eprint
  {http://arxiv.org/abs/1705.00141} {arXiv:1705.00141 [hep-ph]} \BibitemShut
  {NoStop}%
%%CITATION = ARXIV:1705.00141;%%
\bibitem [{\citenamefont {Ali}\ \emph {et~al.}(2017)\citenamefont {Ali},
  \citenamefont {Lange},\ and\ \citenamefont {Stone}}]{Ali:2017jda}%
  \BibitemOpen
  \bibfield  {author} {\bibinfo {author} {\bibfnamefont {A.}~\bibnamefont
  {Ali}}, \bibinfo {author} {\bibfnamefont {J.}~\bibnamefont {Lange}}, \ and\
  \bibinfo {author} {\bibfnamefont {S.}~\bibnamefont {Stone}},\ }\href
  {\doibase 10.1016/j.ppnp.2017.08.003} {\bibfield  {journal} {\bibinfo
  {journal} {Prog.\ Part.\ Nucl.\ Phys.}\ }\textbf {\bibinfo {volume} {{\bf
  97}}},\ \bibinfo {pages} {123} (\bibinfo {year} {2017})},\ \Eprint
  {http://arxiv.org/abs/1706.00610} {arXiv:1706.00610 [hep-ph]} \BibitemShut
  {NoStop}%
%%CITATION = ARXIV:1706.00610;%%
\bibitem [{\citenamefont {Olsen}\ \emph {et~al.}(2018)\citenamefont {Olsen},
  \citenamefont {Skwarnicki},\ and\ \citenamefont {Zieminska}}]{Olsen:2017bmm}%
  \BibitemOpen
  \bibfield  {author} {\bibinfo {author} {\bibfnamefont {S.}~\bibnamefont
  {Olsen}}, \bibinfo {author} {\bibfnamefont {T.}~\bibnamefont {Skwarnicki}}, \
  and\ \bibinfo {author} {\bibfnamefont {D.}~\bibnamefont {Zieminska}},\ }\href
  {\doibase 10.1103/RevModPhys.90.015003} {\bibfield  {journal} {\bibinfo
  {journal} {Rev.\ Mod.\ Phys.}\ }\textbf {\bibinfo {volume} {{\bf 90}}},\
  \bibinfo {pages} {015003} (\bibinfo {year} {2018})},\ \Eprint
  {http://arxiv.org/abs/1708.04012} {arXiv:1708.04012 [hep-ph]} \BibitemShut
  {NoStop}%
%%CITATION = ARXIV:1708.04012;%%
\bibitem [{\citenamefont {Karliner}\ \emph {et~al.}(2018)\citenamefont
  {Karliner}, \citenamefont {Rosner},\ and\ \citenamefont
  {Skwarnicki}}]{Karliner:2017qhf}%
  \BibitemOpen
  \bibfield  {author} {\bibinfo {author} {\bibfnamefont {M.}~\bibnamefont
  {Karliner}}, \bibinfo {author} {\bibfnamefont {J.}~\bibnamefont {Rosner}}, \
  and\ \bibinfo {author} {\bibfnamefont {T.}~\bibnamefont {Skwarnicki}},\
  }\href {\doibase 10.1146/annurev-nucl-101917-020902} {\bibfield  {journal}
  {\bibinfo  {journal} {Annu.\ Rev.\ Nucl.\ Part.\ Sci.}\ }\textbf {\bibinfo
  {volume} {{\bf 68}}},\ \bibinfo {pages} {17} (\bibinfo {year} {2018})},\
  \Eprint {http://arxiv.org/abs/1711.10626} {arXiv:1711.10626 [hep-ph]}
  \BibitemShut {NoStop}%
%%CITATION = ARXIV:1711.10626;%%
\bibitem [{\citenamefont {Yuan}(2018)}]{Yuan:2018inv}%
  \BibitemOpen
  \bibfield  {author} {\bibinfo {author} {\bibfnamefont {C.-Z.}\ \bibnamefont
  {Yuan}},\ }\href {\doibase 10.1142/S0217751X18300181} {\bibfield  {journal}
  {\bibinfo  {journal} {Int.\ J. Mod.\ Phys.\ A}\ }\textbf {\bibinfo {volume}
  {{\bf 33}}},\ \bibinfo {pages} {1830018} (\bibinfo {year} {2018})},\ \Eprint
  {http://arxiv.org/abs/1808.01570} {arXiv:1808.01570 [hep-ex]} \BibitemShut
  {NoStop}%
%%CITATION = ARXIV:1808.01570;%%
\bibitem [{\citenamefont {Liu}\ \emph {et~al.}(2019)\citenamefont {Liu},
  \citenamefont {Chen}, \citenamefont {Chen}, \citenamefont {Liu},\ and\
  \citenamefont {Zhu}}]{Liu:2019zoy}%
  \BibitemOpen
  \bibfield  {author} {\bibinfo {author} {\bibfnamefont {Y.-R.}\ \bibnamefont
  {Liu}}, \bibinfo {author} {\bibfnamefont {H.-X.}\ \bibnamefont {Chen}},
  \bibinfo {author} {\bibfnamefont {W.}~\bibnamefont {Chen}}, \bibinfo {author}
  {\bibfnamefont {X.}~\bibnamefont {Liu}}, \ and\ \bibinfo {author}
  {\bibfnamefont {S.-L.}\ \bibnamefont {Zhu}},\ }\href {\doibase
  10.1016/j.ppnp.2019.04.003} {\bibfield  {journal} {\bibinfo  {journal}
  {Prog.\ Part.\ Nucl.\ Phys.}\ }\textbf {\bibinfo {volume} {{\bf 107}}},\
  \bibinfo {pages} {237} (\bibinfo {year} {2019})},\ \Eprint
  {http://arxiv.org/abs/1903.11976} {arXiv:1903.11976 [hep-ph]} \BibitemShut
  {NoStop}%
%%CITATION = ARXIV:1903.11976;%%
\bibitem [{\citenamefont {Brambilla}\ \emph {et~al.}(2020)\citenamefont
  {Brambilla}, \citenamefont {Eidelman}, \citenamefont {Hanhart}, \citenamefont
  {Nefediev}, \citenamefont {Shen}, \citenamefont {Thomas}, \citenamefont
  {Vairo},\ and\ \citenamefont {Yuan}}]{Brambilla:2019esw}%
  \BibitemOpen
  \bibfield  {author} {\bibinfo {author} {\bibfnamefont {N.}~\bibnamefont
  {Brambilla}}, \bibinfo {author} {\bibfnamefont {S.}~\bibnamefont {Eidelman}},
  \bibinfo {author} {\bibfnamefont {C.}~\bibnamefont {Hanhart}}, \bibinfo
  {author} {\bibfnamefont {A.}~\bibnamefont {Nefediev}}, \bibinfo {author}
  {\bibfnamefont {C.-P.}\ \bibnamefont {Shen}}, \bibinfo {author}
  {\bibfnamefont {C.}~\bibnamefont {Thomas}}, \bibinfo {author} {\bibfnamefont
  {A.}~\bibnamefont {Vairo}}, \ and\ \bibinfo {author} {\bibfnamefont {C.-Z.}\
  \bibnamefont {Yuan}},\ }\href {\doibase 10.1016/j.physrep.2020.05.001}
  {\bibfield  {journal} {\bibinfo  {journal} {Phys.\ Rept.}\ }\textbf {\bibinfo
  {volume} {{\bf 873}}},\ \bibinfo {pages} {1} (\bibinfo {year} {2020})},\
  \Eprint {http://arxiv.org/abs/1907.07583} {arXiv:1907.07583 [hep-ex]}
  \BibitemShut {NoStop}%
\bibitem [{\citenamefont {Brodsky}\ \emph {et~al.}(2014)\citenamefont
  {Brodsky}, \citenamefont {Hwang},\ and\ \citenamefont
  {Lebed}}]{Brodsky:2014xia}%
  \BibitemOpen
  \bibfield  {author} {\bibinfo {author} {\bibfnamefont {S.}~\bibnamefont
  {Brodsky}}, \bibinfo {author} {\bibfnamefont {D.}~\bibnamefont {Hwang}}, \
  and\ \bibinfo {author} {\bibfnamefont {R.}~\bibnamefont {Lebed}},\ }\href
  {\doibase 10.1103/PhysRevLett.113.112001} {\bibfield  {journal} {\bibinfo
  {journal} {Phys.\ Rev.\ Lett.}\ }\textbf {\bibinfo {volume} {{\bf 113}}},\
  \bibinfo {pages} {112001} (\bibinfo {year} {2014})},\ \Eprint
  {http://arxiv.org/abs/1406.7281} {arXiv:1406.7281 [hep-ph]} \BibitemShut
  {NoStop}%
%%CITATION = ARXIV:1406.7281;%%
\bibitem [{\citenamefont {Lebed}(2015)}]{Lebed:2015tna}%
  \BibitemOpen
  \bibfield  {author} {\bibinfo {author} {\bibfnamefont {R.}~\bibnamefont
  {Lebed}},\ }\href {\doibase 10.1016/j.physletb.2015.08.032} {\bibfield
  {journal} {\bibinfo  {journal} {Phys.\ Lett.\ B}\ }\textbf {\bibinfo {volume}
  {{\bf 749}}},\ \bibinfo {pages} {454} (\bibinfo {year} {2015})},\ \Eprint
  {http://arxiv.org/abs/1507.05867} {arXiv:1507.05867 [hep-ph]} \BibitemShut
  {NoStop}%
%%CITATION = ARXIV:1507.05867;%%
\bibitem [{\citenamefont {Lebed}(2017)}]{Lebed:2017min}%
  \BibitemOpen
  \bibfield  {author} {\bibinfo {author} {\bibfnamefont {R.}~\bibnamefont
  {Lebed}},\ }\href {\doibase 10.1103/PhysRevD.96.116003} {\bibfield  {journal}
  {\bibinfo  {journal} {Phys.\ Rev.\ D}\ }\textbf {\bibinfo {volume} {{\bf
  96}}},\ \bibinfo {pages} {116003} (\bibinfo {year} {2017})},\ \Eprint
  {http://arxiv.org/abs/1709.06097} {arXiv:1709.06097 [hep-ph]} \BibitemShut
  {NoStop}%
%%CITATION = ARXIV:1709.06097;%%
\bibitem [{\citenamefont {Juge}\ \emph {et~al.}(1998)\citenamefont {Juge},
  \citenamefont {Kuti},\ and\ \citenamefont {Morningstar}}]{Juge:1997nc}%
  \BibitemOpen
  \bibfield  {author} {\bibinfo {author} {\bibfnamefont {K.}~\bibnamefont
  {Juge}}, \bibinfo {author} {\bibfnamefont {J.}~\bibnamefont {Kuti}}, \ and\
  \bibinfo {author} {\bibfnamefont {C.}~\bibnamefont {Morningstar}},\
  }\bibfield  {booktitle} {\emph {\bibinfo {booktitle} {{Contents of LAT97
  proceedings}}},\ }\href {\doibase 10.1016/S0920-5632(97)00759-7} {\bibfield
  {journal} {\bibinfo  {journal} {Nucl.\ Phys.\ Proc.\ Suppl.}\ }\textbf
  {\bibinfo {volume} {{\bf 63}}},\ \bibinfo {pages} {326} (\bibinfo {year}
  {1998})},\ \Eprint {http://arxiv.org/abs/hep-lat/9709131}
  {arXiv:hep-lat/9709131 [hep-lat]} \BibitemShut {NoStop}%
%%CITATION = HEP-LAT/9709131;%%
\bibitem [{\citenamefont {Juge}\ \emph {et~al.}(1999)\citenamefont {Juge},
  \citenamefont {Kuti},\ and\ \citenamefont {Morningstar}}]{Juge:1999ie}%
  \BibitemOpen
  \bibfield  {author} {\bibinfo {author} {\bibfnamefont {K.}~\bibnamefont
  {Juge}}, \bibinfo {author} {\bibfnamefont {J.}~\bibnamefont {Kuti}}, \ and\
  \bibinfo {author} {\bibfnamefont {C.}~\bibnamefont {Morningstar}},\ }\href
  {\doibase 10.1103/PhysRevLett.82.4400} {\bibfield  {journal} {\bibinfo
  {journal} {Phys.\ Rev.\ Lett.}\ }\textbf {\bibinfo {volume} {{\bf 82}}},\
  \bibinfo {pages} {4400} (\bibinfo {year} {1999})},\ \Eprint
  {http://arxiv.org/abs/hep-ph/9902336} {arXiv:hep-ph/9902336 [hep-ph]}
  \BibitemShut {NoStop}%
%%CITATION = HEP-PH/9902336;%%
\bibitem [{\citenamefont {Juge}\ \emph {et~al.}(2003)\citenamefont {Juge},
  \citenamefont {Kuti},\ and\ \citenamefont {Morningstar}}]{Juge:2002br}%
  \BibitemOpen
  \bibfield  {author} {\bibinfo {author} {\bibfnamefont {K.}~\bibnamefont
  {Juge}}, \bibinfo {author} {\bibfnamefont {J.}~\bibnamefont {Kuti}}, \ and\
  \bibinfo {author} {\bibfnamefont {C.}~\bibnamefont {Morningstar}},\ }\href
  {\doibase 10.1103/PhysRevLett.90.161601} {\bibfield  {journal} {\bibinfo
  {journal} {Phys.\ Rev.\ Lett.}\ }\textbf {\bibinfo {volume} {{\bf 90}}},\
  \bibinfo {pages} {161601} (\bibinfo {year} {2003})},\ \Eprint
  {http://arxiv.org/abs/hep-lat/0207004} {arXiv:hep-lat/0207004 [hep-lat]}
  \BibitemShut {NoStop}%
%%CITATION = HEP-LAT/0207004;%%
\bibitem [{Mor()}]{Morningstar:2019}%
  \BibitemOpen
  \href@noop {} {}\bibinfo {howpublished}
  {\url{http://www.andrew.cmu.edu/user/cmorning/static_potentials/SU3_4D/greet.html}}\BibitemShut
  {NoStop}%
\bibitem [{\citenamefont {Capitani}\ \emph {et~al.}(2019)\citenamefont
  {Capitani}, \citenamefont {Philipsen}, \citenamefont {Reisinger},
  \citenamefont {Riehl},\ and\ \citenamefont {Wagner}}]{Capitani:2018rox}%
  \BibitemOpen
  \bibfield  {author} {\bibinfo {author} {\bibfnamefont {S.}~\bibnamefont
  {Capitani}}, \bibinfo {author} {\bibfnamefont {O.}~\bibnamefont {Philipsen}},
  \bibinfo {author} {\bibfnamefont {C.}~\bibnamefont {Reisinger}}, \bibinfo
  {author} {\bibfnamefont {C.}~\bibnamefont {Riehl}}, \ and\ \bibinfo {author}
  {\bibfnamefont {M.}~\bibnamefont {Wagner}},\ }\href {\doibase
  10.1103/PhysRevD.99.034502} {\bibfield  {journal} {\bibinfo  {journal}
  {Phys.\ Rev.\ D}\ }\textbf {\bibinfo {volume} {{\bf 99}}},\ \bibinfo {pages}
  {034502} (\bibinfo {year} {2019})},\ \Eprint
  {http://arxiv.org/abs/1811.11046} {arXiv:1811.11046 [hep-lat]} \BibitemShut
  {NoStop}%
%%CITATION = ARXIV:1811.11046;%%
\bibitem [{\citenamefont {Giron}\ \emph {et~al.}(2019)\citenamefont {Giron},
  \citenamefont {Lebed},\ and\ \citenamefont {Peterson}}]{Giron:2019bcs}%
  \BibitemOpen
  \bibfield  {author} {\bibinfo {author} {\bibfnamefont {J.}~\bibnamefont
  {Giron}}, \bibinfo {author} {\bibfnamefont {R.}~\bibnamefont {Lebed}}, \ and\
  \bibinfo {author} {\bibfnamefont {C.}~\bibnamefont {Peterson}},\ }\href
  {\doibase 10.1007/JHEP05(2019)061} {\bibfield  {journal} {\bibinfo  {journal}
  {J. High Energy Phys.}\ }\textbf {\bibinfo {volume} {{\bf 05}}},\ \bibinfo
  {pages} {061} (\bibinfo {year} {2019})},\ \Eprint
  {http://arxiv.org/abs/1903.04551} {arXiv:1903.04551 [hep-ph]} \BibitemShut
  {NoStop}%
%%CITATION = ARXIV:1903.04551;%%
\bibitem [{\citenamefont {Giron}\ \emph {et~al.}(2020)\citenamefont {Giron},
  \citenamefont {Lebed},\ and\ \citenamefont {Peterson}}]{Giron:2019cfc}%
  \BibitemOpen
  \bibfield  {author} {\bibinfo {author} {\bibfnamefont {J.}~\bibnamefont
  {Giron}}, \bibinfo {author} {\bibfnamefont {R.}~\bibnamefont {Lebed}}, \ and\
  \bibinfo {author} {\bibfnamefont {C.}~\bibnamefont {Peterson}},\ }\href
  {\doibase 10.1007/JHEP01(2020)124} {\bibfield  {journal} {\bibinfo  {journal}
  {J. High Energy Phys.}\ }\textbf {\bibinfo {volume} {{\bf 01}}},\ \bibinfo
  {pages} {124} (\bibinfo {year} {2020})},\ \Eprint
  {http://arxiv.org/abs/1907.08546} {arXiv:1907.08546 [hep-ph]} \BibitemShut
  {NoStop}%
%%CITATION = ARXIV:1907.08546;%%
\bibitem [{\citenamefont {Ghalenovi}\ and\ \citenamefont
  {Sorkhi}(2020)}]{Ghalenovi:2020zen}%
  \BibitemOpen
  \bibfield  {author} {\bibinfo {author} {\bibfnamefont {Z.}~\bibnamefont
  {Ghalenovi}}\ and\ \bibinfo {author} {\bibfnamefont {M.~M.}\ \bibnamefont
  {Sorkhi}},\ }\href {\doibase 10.1140/epjp/s13360-020-00343-6} {\bibfield
  {journal} {\bibinfo  {journal} {Eur. Phys. J. Plus}\ }\textbf {\bibinfo
  {volume} {135}},\ \bibinfo {pages} {399} (\bibinfo {year}
  {2020})}\BibitemShut {NoStop}%
\bibitem [{\citenamefont {Giron}\ and\ \citenamefont
  {Lebed}(2020{\natexlab{a}})}]{Giron:2020fvd}%
  \BibitemOpen
  \bibfield  {author} {\bibinfo {author} {\bibfnamefont {J.}~\bibnamefont
  {Giron}}\ and\ \bibinfo {author} {\bibfnamefont {R.}~\bibnamefont {Lebed}},\
  }\href {\doibase 10.1103/PhysRevD.101.074032} {\bibfield  {journal} {\bibinfo
   {journal} {Phys.\ Rev.\ D}\ }\textbf {\bibinfo {volume} {{\bf 101}}},\
  \bibinfo {pages} {074032} (\bibinfo {year} {2020}{\natexlab{a}})},\ \Eprint
  {http://arxiv.org/abs/2003.02802} {arXiv:2003.02802 [hep-ph]} \BibitemShut
  {NoStop}%
\bibitem [{\citenamefont {Ali}\ \emph {et~al.}(2018)\citenamefont {Ali},
  \citenamefont {Maiani}, \citenamefont {Borisov}, \citenamefont {Ahmed},
  \citenamefont {Jamil~Aslam}, \citenamefont {Parkhomenko}, \citenamefont
  {Polosa},\ and\ \citenamefont {Rehman}}]{Ali:2017wsf}%
  \BibitemOpen
  \bibfield  {author} {\bibinfo {author} {\bibfnamefont {A.}~\bibnamefont
  {Ali}}, \bibinfo {author} {\bibfnamefont {L.}~\bibnamefont {Maiani}},
  \bibinfo {author} {\bibfnamefont {A.}~\bibnamefont {Borisov}}, \bibinfo
  {author} {\bibfnamefont {I.}~\bibnamefont {Ahmed}}, \bibinfo {author}
  {\bibfnamefont {M.}~\bibnamefont {Jamil~Aslam}}, \bibinfo {author}
  {\bibfnamefont {A.}~\bibnamefont {Parkhomenko}}, \bibinfo {author}
  {\bibfnamefont {A.}~\bibnamefont {Polosa}}, \ and\ \bibinfo {author}
  {\bibfnamefont {A.}~\bibnamefont {Rehman}},\ }\href {\doibase
  10.1140/epjc/s10052-017-5501-6} {\bibfield  {journal} {\bibinfo  {journal}
  {Eur.\ Phys.\ J. C}\ }\textbf {\bibinfo {volume} {{\bf 78}}},\ \bibinfo
  {pages} {29} (\bibinfo {year} {2018})},\ \Eprint
  {http://arxiv.org/abs/1708.04650} {arXiv:1708.04650 [hep-ph]} \BibitemShut
  {NoStop}%
\bibitem [{\citenamefont {Giron}\ and\ \citenamefont
  {Lebed}(2020{\natexlab{b}})}]{Giron:2020qpb}%
  \BibitemOpen
  \bibfield  {author} {\bibinfo {author} {\bibfnamefont {J.}~\bibnamefont
  {Giron}}\ and\ \bibinfo {author} {\bibfnamefont {R.}~\bibnamefont {Lebed}},\
  }\href {\doibase 10.1103/PhysRevD.102.014036} {\bibfield  {journal} {\bibinfo
   {journal} {Phys.\ Rev.\ D}\ }\textbf {\bibinfo {volume} {{\bf 102}}},\
  \bibinfo {pages} {014036} (\bibinfo {year} {2020}{\natexlab{b}})},\ \Eprint
  {http://arxiv.org/abs/2005.07100} {arXiv:2005.07100 [hep-ph]} \BibitemShut
  {NoStop}%
\bibitem [{\citenamefont {Lebed}\ and\ \citenamefont
  {Polosa}(2016)}]{Lebed:2016yvr}%
  \BibitemOpen
  \bibfield  {author} {\bibinfo {author} {\bibfnamefont {R.}~\bibnamefont
  {Lebed}}\ and\ \bibinfo {author} {\bibfnamefont {A.}~\bibnamefont {Polosa}},\
  }\href {\doibase 10.1103/PhysRevD.93.094024} {\bibfield  {journal} {\bibinfo
  {journal} {Phys.\ Rev.\ D}\ }\textbf {\bibinfo {volume} {{\bf 93}}},\
  \bibinfo {pages} {094024} (\bibinfo {year} {2016})},\ \Eprint
  {http://arxiv.org/abs/1602.08421} {arXiv:1602.08421 [hep-ph]} \BibitemShut
  {NoStop}%
%%CITATION = ARXIV:1602.08421;%%
\bibitem [{\citenamefont {Shen}\ \emph {et~al.}(2010)\citenamefont {Shen} \emph
  {et~al.}}]{Shen:2009vs}%
  \BibitemOpen
  \bibfield  {author} {\bibinfo {author} {\bibfnamefont {C.}~\bibnamefont
  {Shen}} \emph {et~al.} (\bibinfo {collaboration} {Belle Collaboration}),\
  }\href {\doibase 10.1103/PhysRevLett.104.112004} {\bibfield  {journal}
  {\bibinfo  {journal} {Phys.\ Rev.\ Lett.}\ }\textbf {\bibinfo {volume} {{\bf
  104}}},\ \bibinfo {pages} {112004} (\bibinfo {year} {2010})},\ \Eprint
  {http://arxiv.org/abs/0912.2383} {arXiv:0912.2383 [hep-ex]} \BibitemShut
  {NoStop}%
\bibitem [{\citenamefont {Aaij}\ \emph {et~al.}(2020)\citenamefont {Aaij} \emph
  {et~al.}}]{Aaij:2020fnh}%
  \BibitemOpen
  \bibfield  {author} {\bibinfo {author} {\bibfnamefont {R.}~\bibnamefont
  {Aaij}} \emph {et~al.} (\bibinfo {collaboration} {LHCb Collaboration}),\
  }\href {\doibase 10.1016/j.scib.2020.08.032} {\bibfield  {journal} {\bibinfo
  {journal} {Sci.\ Bull.}\ }\textbf {\bibinfo {volume} {{\bf 65}}},\ \bibinfo
  {pages} {1983} (\bibinfo {year} {2020})},\ \Eprint
  {http://arxiv.org/abs/2006.16957} {arXiv:2006.16957 [hep-ex]} \BibitemShut
  {NoStop}%
\bibitem [{\citenamefont {Giron}\ and\ \citenamefont
  {Lebed}(2020{\natexlab{c}})}]{Giron:2020wpx}%
  \BibitemOpen
  \bibfield  {author} {\bibinfo {author} {\bibfnamefont {J.}~\bibnamefont
  {Giron}}\ and\ \bibinfo {author} {\bibfnamefont {R.}~\bibnamefont {Lebed}},\
  }\href {\doibase 10.1103/PhysRevD.102.074003} {\bibfield  {journal} {\bibinfo
   {journal} {Phys.\ Rev.\ D}\ }\textbf {\bibinfo {volume} {{\bf 102}}},\
  \bibinfo {pages} {074003} (\bibinfo {year} {2020}{\natexlab{c}})},\ \Eprint
  {http://arxiv.org/abs/2008.01631} {arXiv:2008.01631 [hep-ph]} \BibitemShut
  {NoStop}%
\bibitem [{\citenamefont {Maiani}\ \emph {et~al.}(2014)\citenamefont {Maiani},
  \citenamefont {Piccinini}, \citenamefont {Polosa},\ and\ \citenamefont
  {Riquer}}]{Maiani:2014aja}%
  \BibitemOpen
  \bibfield  {author} {\bibinfo {author} {\bibfnamefont {L.}~\bibnamefont
  {Maiani}}, \bibinfo {author} {\bibfnamefont {F.}~\bibnamefont {Piccinini}},
  \bibinfo {author} {\bibfnamefont {A.}~\bibnamefont {Polosa}}, \ and\ \bibinfo
  {author} {\bibfnamefont {V.}~\bibnamefont {Riquer}},\ }\href {\doibase
  10.1103/PhysRevD.89.114010} {\bibfield  {journal} {\bibinfo  {journal}
  {Phys.\ Rev.\ D}\ }\textbf {\bibinfo {volume} {{\bf 89}}},\ \bibinfo {pages}
  {114010} (\bibinfo {year} {2014})},\ \Eprint {http://arxiv.org/abs/1405.1551}
  {arXiv:1405.1551 [hep-ph]} \BibitemShut {NoStop}%
%%CITATION = ARXIV:1405.1551;%%
\bibitem [{\citenamefont {Chen}\ \emph {et~al.}(2015)\citenamefont {Chen},
  \citenamefont {Maiani}, \citenamefont {Polosa},\ and\ \citenamefont
  {Riquer}}]{Chen:2015dig}%
  \BibitemOpen
  \bibfield  {author} {\bibinfo {author} {\bibfnamefont {H.-X.}\ \bibnamefont
  {Chen}}, \bibinfo {author} {\bibfnamefont {L.}~\bibnamefont {Maiani}},
  \bibinfo {author} {\bibfnamefont {A.}~\bibnamefont {Polosa}}, \ and\ \bibinfo
  {author} {\bibfnamefont {V.}~\bibnamefont {Riquer}},\ }\href {\doibase
  10.1140/epjc/s10052-015-3781-2} {\bibfield  {journal} {\bibinfo  {journal}
  {Eur. Phys. J. C}\ }\textbf {\bibinfo {volume} {75}},\ \bibinfo {pages} {550}
  (\bibinfo {year} {2015})},\ \Eprint {http://arxiv.org/abs/1510.03626}
  {arXiv:1510.03626 [hep-ph]} \BibitemShut {NoStop}%
\bibitem [{\citenamefont {Ablikim}\ \emph {et~al.}(2014)\citenamefont {Ablikim}
  \emph {et~al.}}]{Ablikim:2013dyn}%
  \BibitemOpen
  \bibfield  {author} {\bibinfo {author} {\bibfnamefont {M.}~\bibnamefont
  {Ablikim}} \emph {et~al.} (\bibinfo {collaboration} {BESIII Collaboration}),\
  }\href {\doibase 10.1103/PhysRevLett.112.092001} {\bibfield  {journal}
  {\bibinfo  {journal} {Phys. Rev. Lett.}\ }\textbf {\bibinfo {volume} {112}},\
  \bibinfo {pages} {092001} (\bibinfo {year} {2014})},\ \Eprint
  {http://arxiv.org/abs/1310.4101} {arXiv:1310.4101 [hep-ex]} \BibitemShut
  {NoStop}%
\bibitem [{\citenamefont {Ablikim}\ \emph
  {et~al.}(2019{\natexlab{a}})\citenamefont {Ablikim} \emph
  {et~al.}}]{Ablikim:2019zio}%
  \BibitemOpen
  \bibfield  {author} {\bibinfo {author} {\bibfnamefont {M.}~\bibnamefont
  {Ablikim}} \emph {et~al.} (\bibinfo {collaboration} {BESIII Collaboration}),\
  }\href {\doibase 10.1103/PhysRevLett.122.232002} {\bibfield  {journal}
  {\bibinfo  {journal} {Phys.\ Rev.\ Lett.}\ }\textbf {\bibinfo {volume} {{\bf
  122}}},\ \bibinfo {pages} {232002} (\bibinfo {year} {2019}{\natexlab{a}})},\
  \Eprint {http://arxiv.org/abs/1903.04695} {arXiv:1903.04695 [hep-ex]}
  \BibitemShut {NoStop}%
%%CITATION = ARXIV:1903.04695;%%
\bibitem [{\citenamefont {Ablikim}\ \emph {et~al.}(2013)\citenamefont {Ablikim}
  \emph {et~al.}}]{Ablikim:2013mio}%
  \BibitemOpen
  \bibfield  {author} {\bibinfo {author} {\bibfnamefont {M.}~\bibnamefont
  {Ablikim}} \emph {et~al.} (\bibinfo {collaboration} {BESIII Collaboration}),\
  }\href {\doibase 10.1103/PhysRevLett.110.252001} {\bibfield  {journal}
  {\bibinfo  {journal} {Phys. Rev. Lett.}\ }\textbf {\bibinfo {volume} {110}},\
  \bibinfo {pages} {252001} (\bibinfo {year} {2013})},\ \Eprint
  {http://arxiv.org/abs/1303.5949} {arXiv:1303.5949 [hep-ex]} \BibitemShut
  {NoStop}%
\bibitem [{\citenamefont {Liu}\ \emph {et~al.}(2013)\citenamefont {Liu} \emph
  {et~al.}}]{Liu:2013dau}%
  \BibitemOpen
  \bibfield  {author} {\bibinfo {author} {\bibfnamefont {Z.}~\bibnamefont
  {Liu}} \emph {et~al.} (\bibinfo {collaboration} {Belle Collaboration}),\
  }\href {\doibase 10.1103/PhysRevLett.110.252002} {\bibfield  {journal}
  {\bibinfo  {journal} {Phys. Rev. Lett.}\ }\textbf {\bibinfo {volume} {110}},\
  \bibinfo {pages} {252002} (\bibinfo {year} {2013})},\ \bibinfo {note}
  {[Erratum: Phys.\ Rev.\ Lett.\ {\bf 111}, 019901 (2013)]},\ \Eprint
  {http://arxiv.org/abs/1304.0121} {arXiv:1304.0121 [hep-ex]} \BibitemShut
  {NoStop}%
\bibitem [{\citenamefont {Ablikim}\ \emph {et~al.}(2020)\citenamefont {Ablikim}
  \emph {et~al.}}]{BESIII:2020pov}%
  \BibitemOpen
  \bibfield  {author} {\bibinfo {author} {\bibfnamefont {M.}~\bibnamefont
  {Ablikim}} \emph {et~al.} (\bibinfo {collaboration} {BESIII Collaboration}),\
  }\href {\doibase 10.1103/PhysRevD.102.012009} {\bibfield  {journal} {\bibinfo
   {journal} {Phys.\ Rev.\ D}\ }\textbf {\bibinfo {volume} {{\bf 102}}},\
  \bibinfo {pages} {012009} (\bibinfo {year} {2020})},\ \Eprint
  {http://arxiv.org/abs/2004.13788} {arXiv:2004.13788 [hep-ex]} \BibitemShut
  {NoStop}%
\bibitem [{\citenamefont {Ablikim}\ \emph
  {et~al.}(2017{\natexlab{a}})\citenamefont {Ablikim} \emph
  {et~al.}}]{Ablikim:2016qzw}%
  \BibitemOpen
  \bibfield  {author} {\bibinfo {author} {\bibfnamefont {M.}~\bibnamefont
  {Ablikim}} \emph {et~al.} (\bibinfo {collaboration} {BESIII Collaboration}),\
  }\href {\doibase 10.1103/PhysRevLett.118.092001} {\bibfield  {journal}
  {\bibinfo  {journal} {Phys.\ Rev.\ Lett.}\ }\textbf {\bibinfo {volume} {{\bf
  118}}},\ \bibinfo {pages} {092001} (\bibinfo {year} {2017}{\natexlab{a}})},\
  \Eprint {http://arxiv.org/abs/1611.01317} {arXiv:1611.01317 [hep-ex]}
  \BibitemShut {NoStop}%
%%CITATION = ARXIV:1611.01317;%%
\bibitem [{\citenamefont {Aaij}\ \emph
  {et~al.}(2014{\natexlab{a}})\citenamefont {Aaij} \emph
  {et~al.}}]{Aaij:2014ala}%
  \BibitemOpen
  \bibfield  {author} {\bibinfo {author} {\bibfnamefont {R.}~\bibnamefont
  {Aaij}} \emph {et~al.} (\bibinfo {collaboration} {LHCb Collaboration}),\
  }\href {\doibase 10.1016/j.nuclphysb.2014.06.011} {\bibfield  {journal}
  {\bibinfo  {journal} {Nucl.\ Phys.\ B}\ }\textbf {\bibinfo {volume} {{\bf
  886}}},\ \bibinfo {pages} {665} (\bibinfo {year} {2014}{\natexlab{a}})},\
  \Eprint {http://arxiv.org/abs/1404.0275} {arXiv:1404.0275 [hep-ex]}
  \BibitemShut {NoStop}%
\bibitem [{\citenamefont {Ablikim}\ \emph {et~al.}(2021)\citenamefont {Ablikim}
  \emph {et~al.}}]{Ablikim:2021rba}%
  \BibitemOpen
  \bibfield  {author} {\bibinfo {author} {\bibfnamefont {M.}~\bibnamefont
  {Ablikim}} \emph {et~al.} (\bibinfo {collaboration} {BESIII Collaboration}),\
  }\href@noop {} {\  (\bibinfo {year} {2021})},\ \Eprint
  {http://arxiv.org/abs/2102.00644} {arXiv:2102.00644 [hep-ex]} \BibitemShut
  {NoStop}%
\bibitem [{\citenamefont {Brambilla}\ \emph {et~al.}(2004)\citenamefont
  {Brambilla} \emph {et~al.}}]{Brambilla:2004wf}%
  \BibitemOpen
  \bibfield  {author} {\bibinfo {author} {\bibfnamefont {N.}~\bibnamefont
  {Brambilla}} \emph {et~al.} (\bibinfo {collaboration} {Quarkonium Working
  Group}),\ }\href {\doibase 10.5170/CERN-2005-005} {\  (\bibinfo {year}
  {2004}),\ 10.5170/CERN-2005-005},\ \bibinfo {note} {\mbox{Chap.\ 4}, Sec.\ 6.
  written by Eichten, E.},\ \Eprint {http://arxiv.org/abs/hep-ph/0412158}
  {arXiv:hep-ph/0412158} \BibitemShut {NoStop}%
\bibitem [{\citenamefont {Zyla}\ \emph {et~al.}(2020)\citenamefont {Zyla} \emph
  {et~al.}}]{Zyla:2020zbs}%
  \BibitemOpen
  \bibfield  {author} {\bibinfo {author} {\bibfnamefont {P.}~\bibnamefont
  {Zyla}} \emph {et~al.} (\bibinfo {collaboration} {Particle Data Group}),\
  }\href {\doibase 10.1093/ptep/ptaa104} {\bibfield  {journal} {\bibinfo
  {journal} {PTEP}\ }\textbf {\bibinfo {volume} {{\bf 2020}}},\ \bibinfo
  {pages} {083C01} (\bibinfo {year} {2020})}\BibitemShut {NoStop}%
\bibitem [{\citenamefont {de~Shalit}\ and\ \citenamefont
  {Talmi}(1963)}]{Shalit:1963nuclear}%
  \BibitemOpen
  \bibfield  {author} {\bibinfo {author} {\bibfnamefont {A.}~\bibnamefont
  {de~Shalit}}\ and\ \bibinfo {author} {\bibfnamefont {I.}~\bibnamefont
  {Talmi}},\ }\href@noop {} {\emph {\bibinfo {title} {\it Nuclear Shell
  Theory}}},\ Pure and Applied Physics\ (\bibinfo  {publisher} {Academic Press,
  New York},\ \bibinfo {year} {1963})\BibitemShut {NoStop}%
\bibitem [{\citenamefont {Ablikim}\ \emph
  {et~al.}(2019{\natexlab{b}})\citenamefont {Ablikim} \emph
  {et~al.}}]{Ablikim:2018vxx}%
  \BibitemOpen
  \bibfield  {author} {\bibinfo {author} {\bibfnamefont {M.}~\bibnamefont
  {Ablikim}} \emph {et~al.} (\bibinfo {collaboration} {BESIII Collaboration}),\
  }\href {\doibase 10.1103/PhysRevLett.122.102002} {\bibfield  {journal}
  {\bibinfo  {journal} {Phys.\ Rev.\ Lett.}\ }\textbf {\bibinfo {volume}
  {122}},\ \bibinfo {pages} {102002} (\bibinfo {year} {2019}{\natexlab{b}})},\
  \Eprint {http://arxiv.org/abs/1808.02847} {arXiv:1808.02847 [hep-ex]}
  \BibitemShut {NoStop}%
%%CITATION = ARXIV:1808.02847;%%
\bibitem [{\citenamefont {Ablikim}\ \emph
  {et~al.}(2019{\natexlab{c}})\citenamefont {Ablikim} \emph
  {et~al.}}]{Ablikim:2019apl}%
  \BibitemOpen
  \bibfield  {author} {\bibinfo {author} {\bibfnamefont {M.}~\bibnamefont
  {Ablikim}} \emph {et~al.} (\bibinfo {collaboration} {BESIII Collaboration}),\
  }\href {\doibase 10.1103/PhysRevD.99.091103} {\bibfield  {journal} {\bibinfo
  {journal} {Phys.\ Rev.\ D}\ }\textbf {\bibinfo {volume} {{\bf 99}}},\
  \bibinfo {pages} {091103} (\bibinfo {year} {2019}{\natexlab{c}})},\ \Eprint
  {http://arxiv.org/abs/1903.02359} {arXiv:1903.02359 [hep-ex]} \BibitemShut
  {NoStop}%
%%CITATION = ARXIV:1903.02359;%%
\bibitem [{\citenamefont {Ablikim}\ \emph
  {et~al.}(2017{\natexlab{b}})\citenamefont {Ablikim} \emph
  {et~al.}}]{Ablikim:2017oaf}%
  \BibitemOpen
  \bibfield  {author} {\bibinfo {author} {\bibfnamefont {M.}~\bibnamefont
  {Ablikim}} \emph {et~al.} (\bibinfo {collaboration} {BESIII Collaboration}),\
  }\href {\doibase 10.1103/PhysRevD.96.032004} {\bibfield  {journal} {\bibinfo
  {journal} {Phys.\ Rev.\ D}\ }\textbf {\bibinfo {volume} {{\bf 96}}},\
  \bibinfo {pages} {032004} (\bibinfo {year} {2017}{\natexlab{b}})},\ \bibinfo
  {note} {[Erratum: Phys.\ Rev.\ D {\bf 99}, 019903 (2019)]},\ \Eprint
  {http://arxiv.org/abs/1703.08787} {arXiv:1703.08787 [hep-ex]} \BibitemShut
  {NoStop}%
%%CITATION = ARXIV:1703.08787;%%
\bibitem [{\citenamefont {Barnes}\ \emph {et~al.}(2005)\citenamefont {Barnes},
  \citenamefont {Godfrey},\ and\ \citenamefont {Swanson}}]{Barnes:2005pb}%
  \BibitemOpen
  \bibfield  {author} {\bibinfo {author} {\bibfnamefont {T.}~\bibnamefont
  {Barnes}}, \bibinfo {author} {\bibfnamefont {S.}~\bibnamefont {Godfrey}}, \
  and\ \bibinfo {author} {\bibfnamefont {E.}~\bibnamefont {Swanson}},\ }\href
  {\doibase 10.1103/PhysRevD.72.054026} {\bibfield  {journal} {\bibinfo
  {journal} {Phys.\ Rev.\ D}\ }\textbf {\bibinfo {volume} {{\bf 72}}},\
  \bibinfo {pages} {054026} (\bibinfo {year} {2005})},\ \Eprint
  {http://arxiv.org/abs/hep-ph/0505002} {arXiv:hep-ph/0505002 [hep-ph]}
  \BibitemShut {NoStop}%
%%CITATION = HEP-PH/0505002;%%
\bibitem [{\citenamefont {Aaij}\ \emph
  {et~al.}(2014{\natexlab{b}})\citenamefont {Aaij} \emph
  {et~al.}}]{Aaij:2014jqa}%
  \BibitemOpen
  \bibfield  {author} {\bibinfo {author} {\bibfnamefont {R.}~\bibnamefont
  {Aaij}} \emph {et~al.} (\bibinfo {collaboration} {LHCb Collaboration}),\
  }\href {\doibase 10.1103/PhysRevLett.112.222002} {\bibfield  {journal}
  {\bibinfo  {journal} {Phys.\ Rev.\ Lett.}\ }\textbf {\bibinfo {volume}
  {112}},\ \bibinfo {pages} {222002} (\bibinfo {year} {2014}{\natexlab{b}})},\
  \Eprint {http://arxiv.org/abs/1404.1903} {arXiv:1404.1903 [hep-ex]}
  \BibitemShut {NoStop}%
%%CITATION = ARXIV:1404.1903;%%
\bibitem [{\citenamefont {Jia}\ \emph {et~al.}(2019)\citenamefont {Jia} \emph
  {et~al.}}]{Jia:2019gfe}%
  \BibitemOpen
  \bibfield  {author} {\bibinfo {author} {\bibfnamefont {S.}~\bibnamefont
  {Jia}} \emph {et~al.} (\bibinfo {collaboration} {Belle Collaboration}),\
  }\href {\doibase 10.1103/PhysRevD.100.111103} {\bibfield  {journal} {\bibinfo
   {journal} {Phys.\ Rev.\ D}\ }\textbf {\bibinfo {volume} {{\bf 100}}},\
  \bibinfo {pages} {111103} (\bibinfo {year} {2019})},\ \Eprint
  {http://arxiv.org/abs/1911.00671} {arXiv:1911.00671 [hep-ex]} \BibitemShut
  {NoStop}%
%%CITATION = ARXIV:1911.00671;%%
\bibitem [{\citenamefont {Jia}\ \emph {et~al.}(2020)\citenamefont {Jia} \emph
  {et~al.}}]{Jia:2020epr}%
  \BibitemOpen
  \bibfield  {author} {\bibinfo {author} {\bibfnamefont {S.}~\bibnamefont
  {Jia}} \emph {et~al.} (\bibinfo {collaboration} {Belle Collaboration}),\
  }\href {\doibase 10.1103/PhysRevD.101.091101} {\bibfield  {journal} {\bibinfo
   {journal} {Phys.\ Rev.\ D}\ }\textbf {\bibinfo {volume} {{\bf 101}}},\
  \bibinfo {pages} {091101} (\bibinfo {year} {2020})},\ \Eprint
  {http://arxiv.org/abs/2004.02404} {arXiv:2004.02404 [hep-ex]} \BibitemShut
  {NoStop}%
%%CITATION = ARXIV:2004.02404;%%
\bibitem [{\citenamefont {Voloshin}(2019)}]{Voloshin:2019ivc}%
  \BibitemOpen
  \bibfield  {author} {\bibinfo {author} {\bibfnamefont {M.}~\bibnamefont
  {Voloshin}},\ }\href {\doibase 10.1103/PhysRevD.99.054028} {\bibfield
  {journal} {\bibinfo  {journal} {Phys.\ Rev.\ D}\ }\textbf {\bibinfo {volume}
  {{\bf 99}}},\ \bibinfo {pages} {054028} (\bibinfo {year} {2019})},\ \Eprint
  {http://arxiv.org/abs/1902.01281} {arXiv:1902.01281 [hep-ph]} \BibitemShut
  {NoStop}%
\end{thebibliography}%
\end{document}